\documentclass[sigconf]{acmart}

\usepackage{dirtytalk}
\usepackage{graphicx}
\usepackage{algorithm}
\usepackage[noend]{algpseudocode}
\usepackage{framed}
\usepackage{subfigure}
\usepackage{colortbl}
\usepackage{wasysym}
\usepackage{multirow}
\usepackage[inline]{enumitem}

\PassOptionsToPackage{unicode}{hyperref}
\PassOptionsToPackage{naturalnames}{hyperref}

\def\BibTeX{{\rm B\kern-.05em{\sc i\kern-.025em b}\kern-.08emT\kern-.1667em\lower.7ex\hbox{E}\kern-.125emX}}

\copyrightyear{2022}
\acmYear{2022}
\setcopyright{rightsretained}
\acmConference[COMPASS '22]{ACM SIGCAS/SIGCHI Conference on Computing and Sustainable Societies (COMPASS)}{June 29-July 1, 2022}{Seattle, WA, USA}
\acmBooktitle{ACM SIGCAS/SIGCHI Conference on Computing and Sustainable Societies (COMPASS) (COMPASS '22), June 29-July 1, 2022, Seattle, WA, USA}
\acmDOI{10.1145/3530190.3534820}
\acmISBN{978-1-4503-9347-8/22/06}

\begin{document}

\title{Telechain: Bridging Telecom Policy and Blockchain Practice} 

\newcommand{\tsc}[1]{\textsuperscript{#1}} 

\author{
    Sudheesh Singanamalla\tsc{$\dagger$,1,2},
    Apurv Mehra\tsc{1},
    Nishanth Chandran\tsc{1},
    Himanshi Lohchab\tsc{3}, \\
    Seshanuradha Chava\tsc{3},
    Asit Kadayan\tsc{4},
    Sunil Bajpai\tsc{4},
    Kurtis Heimerl\tsc{2}, \\
    Richard Anderson\tsc{2},
    Satya Lokam\tsc{1}
}
\affiliation{
	\institution{\vspace{0.7em}}
    \institution{
    	\tsc{1} Microsoft Research India \hfil
    	\tsc{2} University of Washington \hfil
    	\tsc{3} Tanla Platforms Limited \hfil \\
    	\tsc{4} Telecom Regulatory Authority of India, Government of India
    }
    \country{}
}
\thanks{
	$\dagger$ Sudheesh Singanamalla was a Research Fellow at Microsoft Research India during a significant part of this effort and continued it as a graduate student at the University of Washington.
	\textit{Corresponding Authors: sudheesh@cs.washington.edu, sunilbajpai@gmail.com, satya@microsoft.com}
}

\ccsdesc[300]{Applied computing~Computing in government}
\ccsdesc[500]{Social and professional topics~Governmental regulations}
\ccsdesc[300]{Social and professional topics~Computing / technology policy}

\renewcommand{\shortauthors}{Singanamalla et al.}

%
\begin{abstract}

The use of blockchain in regulatory ecosystems is a promising approach to address challenges of compliance among mutually untrusted entities. In this work, we consider applications of blockchain technologies in telecom regulations. In particular, we address growing concerns around Unsolicited Commercial Communication (UCC aka. \textit{spam}) sent through text messages (SMS) and phone calls in India. Despite several regulatory measures taken to curb the menace of spam it continues to be a nuisance to subscribers while posing challenges to telecom operators and regulators alike.

In this paper, we present a consortium blockchain based architecture to address the problem of UCC in India. Our solution improves subscriber experiences, improves the efficiency of regulatory processes while also positively impacting all stakeholders in the telecom ecosystem. Unlike previous approaches to the problem of UCC, which are all  \textit{ex-post}, our approach to adherence to the regulations is \textit{ex-ante}. The proposal described in this paper is a primary contributor to the revision of regulations concerning UCC and spam by the Telecom Regulatory Authority of India (TRAI). The new regulations published in July 2018 were first of a kind in the world and amended the 2010 Telecom Commercial Communication Customer Preference Regulation (TCCCPR), through mandating the use of a blockchain/distributed ledgers in addressing the UCC problem. 
In this paper, we provide a holistic account of of the projects' evolution from (1) its design and strategy, to  (2) regulatory and policy action, (3) country wide implementation and deployment, and (4) evaluation and impact of the work.
While the scope of the work presented in this paper is in the context of the UCC problem in India, we believe that the approach can be generalized to adopt blockchain based solutions to improve regulatory processes in other contexts and countries. We hope this account will serve as a useful case study for the stakeholders of the telecommunications ecosystem and regulators, and motivate countries across the world facing similar challenges to consider the viability of the technology, be convinced to establish it, continue efforts at addressing active research challenges, and scale the technology from our experiences.



\end{abstract}


%
\maketitle

\section{Introduction}
\label{sec:introduction}

Despite several efforts by telecom operators, Internet businesses, operating system and application developers, regulators, the problem of spam in communication networks (Internet and cellular) continues to be a global challenge. India has seen tremendous growth in the telecom industry with more than a billion active subscribers (1197.87 million)  making it one of the largest wireless markets in the world~\cite{coai.com, traidec2018}. The low costs for short message service (SMS) (0.085 INR/SMS for bulk advertisers or 0.001 USD/SMS~\cite{bulksmspricing}) and calls has made it one of the most cost effective ways to reach potential customers and sell services and products using telemarketing.

More than 30 billion SMSs are sent in India every month which contributes a monthly gross revenue of approximately 2 billion dollars for all the telecom operators in the country. There are 12639 legally active and authorized telemarketing organizations in India with some large telemarketers sending up to a billion SMSs on certain days. While the cost effectiveness of the scale has made it easy for businesses to reach out to their target audience, telemarketing has brought with it a serious invasion of privacy and has become a major irritant to mobile subscribers in addition to possible frauds~\cite{toismsscam}. The issue is rampant across the world with the Federal Communications Commission (FCC) in the United States of America mentioning that more than half the calls received by subscribers in 2019 are spam~\cite{cnnfcc2019, johnoliverfcc2019}. Research efforts have shown that the growth of legitimate bulk messaging traffic in the recent years degraded the performance of spam detectors resulting in increasing number of spam messages passing through the established content based filters~\cite{reaves2016detecting}. Prior research in Pakistan has indicated the growing role of SMS messaging for financial inclusion but highlighted and attempted to address the growing concern of fraudulent SMS~\cite{pervaiz2019assessment}. A growing body of work has studied the role of security in global development, and the various factors influencing security and privacy of people in developing regions, while identifying the important role that  national and international policy could play in influencing and improving the security practices~\cite{ben2011computing, vashistha2018examining}.

\textcolor{gray}{\textit{\say{The SMSs appear like regular text messages sent by banks. However, they are sent by fraudsters and contain a link which installs a malware allowing access to the phone} - Cybercrime investigation officer~\cite{toismsscam}}}

To curb the menace of Unsolicited Commercial Communication (UCC), the Telecom Regulatory Authority of India (TRAI) introduced a regulation in 2010 allowing mobile subscribers to register themselves to a National Do Not Disturb (NDND) registry~\cite{traidndapp, dash2009national}. The NDND registry behaves as a preference portal where the subscribers can block all communication or selectively block communication from a predefined set of seven preference categories (Banking, Real estate, Education, Health etc..,). Currently, the NDND registry, since its inception in 2010, holds preference information of more than 230 million subscribers who have exercised their choice and  opted out from receiving promotional content. Additionally, the government mandated that each telecom service provider (TSP)/carrier in the country implement a \emph{common}  procedure for complaint registration and resolution in their networks. So far, more than 2 million complaints have been registered which resulted in the active block-list of 460000 numbers and disconnection of approximately 1.4 million phone numbers for violations and instances of consumer courts levying fines on telecom service providers~\cite{toicourtviolation}.

Despite the introduction of monetary dis-incentives, preference registration, and  complaint resolution and tracking facilities by TRAI there have been new efforts taken by violators to gather phone numbers of potential subscribers from various data leaks in the country~\cite{truecallerindiatoday, huffingtonpostindialeak, mcdonaldsindialeak}, and reach customers using unauthorized commercial communication channels without regard for the customers' preferences. The implementation of the NDND registry came with its own set of challenges. The first involved all the telecom service providers to "sync" their operational databases with the central database hosted by TRAI. This considerably slowed down the process and resulted in a customer's preference being acknowledged and effective only \emph{seven days} after the registration. Additionally, complaint resolution processes currently take up to seven days from their registration because of the heavy need for collaboration among telecom service providers who are generally reluctant to share subscriber information or telemarketer activity.

Motivated by the economic importance of telemarketing, keeping in mind the problems faced by subscribers and the need to improve collaboration, transparency, and accountability in the entire process, this paper describes an architecture that uses consortium blockchains to record customer preferences and extends it to capturing subscriber consent. The solution tackles challenges around data sharing between telecom service providers and processes to improve operational efficiency of current business processes. These suggestions have been incorporated in the latest release of the regulations for unsolicited commercial communication in India by TRAI in July 2018. In addition to detailing the design of the system, we present the journey towards the  realization of the design, its country wide deployment, and challenges faced. Finally, we evaluate the success of the proposal with one year of real world data since the usage of the implemented system from telecommunications operations in India.

We make the following contributions: 
\begin{enumerate*}[label=(\arabic*)]
    \item detail key business processes, operational practices, and challenges faced by the Indian telecommunications ecosystem [\S\ref{sec:current_practices}],
    \item engage with regulatory bodies and stakeholders; design, propose [\S\ref{sec:proposed_solution}], and implement a viable solution addressing performance, security, and privacy challenges with minimal operational impact~[\S\ref{sec:implementation}] resulting in the TCCCPR'18 regulatory amendment,
    \item identify practical challenges during deployment [\S\ref{sec:deployment}],
    \item evaluate the impact of the deployed system [\S\ref{sec:evaluation}],
    \item lay out practical challenges, and limitations of the currently deployed system [\S\ref{sec:limitations}], and
    \item present a discussion [\S\ref{sec:discussion}] and detail potential future efforts [\S\ref{sec:conclusion}].
\end{enumerate*}
\section{Background \& Related Work}

Previous efforts to tackle spam or forms of unsolicited communication fall under two categories, 1) regulatory initiatives taken by the governments to manage spam and 2) technological initiatives and tools built to handle and manage spam. Below, we look at these initiatives \& solutions proposed in the past.


\subsection{Regulatory Initiatives}

The first major regulatory initiative to curb spam was enacted by the United States Congress as the Telephone Consumer Protection Act of 1991 (TCPA 1991), and Telemarketing and Consumer Fraud and Abuse Prevention Act (TCFAPA 1994) which regulate the use of automatic dialers, robocalls (prerecorded calls) \& introduces the national do not call lists~\cite{sorkin1997unsolicited}. Additionally the TCPA'91 mandates that written consent be taken from the customers and that the customer can only be reached out to if they meet a strict opt-in requirement. This meant that the customers are by default opted out of any commercial communication until an explicit consent is provided. Different states in the United States have laws that affect SMS messaging and sales communication which on violation result in serious fines.

In Europe, the Privacy and Electronic Communications Directive 2002 known as the ePrivacy Directive (ePD) established an opt-in approach where any commercial email can only be sent to an individual with a prior agreement or consent~\cite{eu-58-2002}. In the United States, the CAN-SPAM Act of 2003  mandates that all commercial messages sent to a cell phone need to be disclosed as an advertisement and must provide a way to opt-out from receiving future messages while ensuring that they comply with the national do not call list~\cite{canspamact2003}. Similarly, in the United Kingdom, Schedule 2 of the Data Protection Act of 1998 introduces mandatory opt-in for any commercial communication and allows consumers to sue for compensation for receiving spam messages~\cite{uk-dpa-1998}. 

Compared to the CAN-SPAM act, TCPA, and TCFAPA in the United States aimed towards email and spam in telecommunication networks, EU General Data Protection Regulation (GDPR), and Canada's Anti Spam Law (CASL) include laws with much heavier monetary penalties ranging between 1 million dollars (CASL) per violation to 22 million dollars (20 Million EUR). Despite the long term existence of spam prevention regulations, it has only been recent for businesses in the telecom industry to face extremely heavy penalties. The EU GDPR, since its adoption, has penalized telecommunications operators with 69 penalties across different countries in the EU with companies like Telecom Italia and Vodafone Italia facing the largest fines totaling over 30 Million EUR~\cite{telecomitaliagdpr, garanteitaliagdpr, vodafoneitaliagdpr}. 

The first regulation to tackle unsolicited commercial communication in India was released in 2007 introducing the implementation of the national do not call/disturb (DND) register~\cite{dash2009national}, mandating every telemarketer to respect the preference of the subscribers and introducing a mobile identifier code for telemarketers~\cite{traijun2007}. As a default setting all subscribers were opted in to receiving commercial communication unless explicitly opted out by adding their information to the DND register. The regulation also mandated the registration of telemarketers and introduced monetary dis-incentives to curb the problem of spam in the country.

Current regulations mandated worldwide only provided economic dis-incentives for the telemarketers and resulted in banning telemarketers for continued violations~\cite{lee2005can}. 
These acts and regulations have been criticized for increasing the amount of spam by having merely provided guidelines for spammers to spam legally~\cite{sullivan2003spam, moustakas2005combating}. The CAN-SPAM Act was poorly enforced and the economic dis-incentives were not strong enough a deterrent for spammers resulting in increasing email spam~\cite{lee2005can}. For example, in India and many other countries, violators gamed the system of consent by using fine-print on raffle or lucky draw coupons and additionally used leaked phone numbers from databases. Another reason for higher levels of spam could be the default ``opt-in'' approach taken for advertising in India compared to the default ``opt-out'' in Europe and the US. However, the voice call based spam continues to be a global nuisance. This brings in a need for a system that can capture unique consents and map the consent of the customer with the required business entity while ensuring the ability to revoke them as needed. Suggestions were made to revise the act to include long prison sentences~\cite{doi:10.1177/0887403414562604}. In a response to similar regulations in Malaysia, researchers have prescribed \textit{self-help}, \textit{opt-in} \& \textit{opt-out} as strategies to address the spam problem which are now adopted by various governments as do not call lists \& mandatory requirement of consent~\cite{khong2004problem}. However, our experience presented in this paper supports the statement by prior research efforts that legislation alone will not be able to eliminate spam~\cite{moustakas2005combating}. 

\subsection{Technological Initiatives}

There are a variety of mobile apps available for smartphones which help manage and filter spam from the SMS inbox~\cite{microsoftsmsorganizer, truecallersmscategorizer}. Many of these systems use rule based filters which categorize the message based on their promotional nature of text and flag them for review by the user. The usage of mobile identifier codes for telemarketers ensures that the promotional messages can be filtered by using a rule based approach. Other approaches like \textit{SMSAssassin} use a combination of crowd-sourcing and bayesian learning which can be trained over time to identify spam~\cite{Yadav:2011:SCD:2184489.2184491}.

Additionally there are many third party applications on the Google play store for android which allow users to maintain block-lists and ban specific SMS senders. Narayan \textit{et al.} studied the top twelve SMS spam filtering apps and conclude that most of these apps are ineffective and propose a new 2 level stacked classifier approach to detect and classify spam~\cite{Narayan:2013:CCE:2516760.2516772}. Apple's default SMS application (iMessage) and Google messages both provide the option to block each sender using the settings without the need for any third party applications~\cite{pcworldindia}.

Previous work by Jiang \textit{et al.} tries to analyze SMS spam in large cellular networks and indicate spatial correlations using text clustering tools over reported spam SMS data~\cite{jiang2013understanding}. Similarly other techniques focus on identifying SMS spam during possibly legitimate bulk messaging by senders~\cite{reaves2016detecting}. Systems like Greystar leverage the fact that SMS spammers target random phone numbers and employ statistical models based on their footprints through a grey space (e.g. data only numbers) essentially behaving as \textit{honeypots} to detect spammers and block them out of the networks~\cite{jiang2013greystar}. Such systems have been implemented and used by the telecom operators in India to detect and fine violators using a random set of regularly changing honeypot phone numbers.

Unfortunately, none of the existing technology solutions by themselves have been effective in completely curbing spam. As long as the cost of sending an SMS remains low, there is huge motivation from the spammer to get around the filters and find new ways to continue to reach the consumer. To avoid such a problem, in this paper we propose a proactive approach to prevent spam by bringing together infrastructures to record consent, and preferences of the customer, rather than reactive approaches to spam taken by default messaging applications today. The technological approaches described address the ``user end'', or target a specific stakeholder of the ecosystem, in an effort to solve the problem of spam but does not address the overall challenge it poses at the ecosystem level. In our work, we attempt to address the challenges throughout the ecosystem with existing individual solutions complementing our efforts.


\subsection{Use of Blockchain in Government}

A blockchain allows a set of participants to create a shared and tamper resistant ledger of transactions that are validated by a consensus mechanism among them. Such a network decentralizes trust among mutually untrusting participants by creating transparency and accountability through collective verifiability~\cite{peck2017blockchains}. A permissioned blockchain is one in which the write privileges are granted to a consortium of entities. These entities govern the policies of the blockchain and are responsible to propagate and verify transactions in a network. Such a blockchain could have a public read privilege or maybe granted only to auditing agencies. Hyperledger Fabric~\cite{Androulaki:2018:HFD:3190508.3190538} \& Ethereum consortium~\cite{wood2014ethereum, eeaethereum} are some popular blockchain frameworks that enable creation of consortium blockchains.


Smart contracts are protocols which make contractual clauses partially or fully self executing and self enforceable. Governments have begun to experiment with blockchain for the various strategic, organizational, economic, informational and technological benefits it brings~\cite{OLNES2017355, peck2017blockchains}. For example storing land records in distributed ledgers prevents manipulation and reduces corruption ~\cite{kshetri2017}. The government of Estonia, has implemented an X-Road system using blockchain transparently and easily allowing citizens to access their data and control data access~\cite{economistblockchain}.

To effectively leverage the advantages of blockchains they need to be guided by \emph{governance}. In this paper, we describe an approach where the regulatory bodies embraced blockchain as a technology to combat the problem of unsolicited commercial communication but eventually saw possibilities to govern other areas with the same technology.

We believe that UCC has unfortunately acquired a negative connotation as spam but instead can be viewed as a problem of bridging the divergence between ex-ante expectations of regulators and telemarketers and the ex-post perceptions by the recipients (subscribers). Spam is therefore very subjective since its a nuisance for those who explicitly choose to exercise their option to not be disturbed but could also take shape of graySMS. GraySMS, like graymail is a situation where an originally interested consumer who opted into receiving the messages, eventually is no longer interested and marks the same message as spam.
\section{Understanding Operational Practices in Indian Telecom Industry}
\label{sec:current_practices}

India is the world's second largest telecommunications market with over a Billion active subscribers and has the lowest call, message, and Internet data pricing in the world enabled by nation scale hyper-giant telecom operators. In this section, we present the detailed view of the telecom ecosystem prior to our intervention, and the operational practices put in place then for addressing the challenges of UCC.

\subsection{Stakeholders in the Telecom Ecosystem}
\label{ssec:stakeholders}

The telecom ecosystem essentially comprises of the following entities:

\begin{enumerate}
    \item \textit{Telecom operators} are telecom service providers who provide telephony and access to data services.
    \item \textit{Telemarketers} are marketing agencies or personnel who solicit customers through phone calls, SMSs \& auto-dialers either directly or on behalf of an organization which shares their (potential) customer details with the telemarketing organization. 
    \item \textit{Principal entities} are organizations which generate content and leverage telemarketers to reach out to potential customers. These are businesses which use telecommunication services to interact with subscribers.
    \item \textit{Subscribers} are the mobile consumers who use the network services provided by telecom operators and are potential customers to the principal entities.
    \item \textit{Regulatory authorities} are typically government entities that make policies and enforce incentive structures that ensure fairness among participants.
\end{enumerate}

\subsection{History of Regulatory Amendments to UCC in India}

Since the effect of the Telecom Commercial Communications Customer Preference Regulations (TCCCPR) in 2010, TRAI in the 6\textsuperscript{th} amendment introduced a SMS header format (AB-XXXXXX) where \textit{AB} corresponds to the telecom provider and region and \textit{XXXXXX} corresponds to the alpha numeric code for the business entity sending the promotional SMS. The amendment to the regulation clearly defined \textit{promotional} \& \textit{transactional} messages in addition to demarcating special phone number sequences which begin with \textit{+91-140} for telemarketers to use during promotional phone calls~\cite{traireg2010}. A \textit{promotional} SMS is used to send offers, discounts or promotions to new or existing customers which may have not been solicited by the recipients and can only be sent between 9AM and 9PM. A \textit{transactional} SMS is used to send one time passwords, informational messages, booking information or order alerts to registered customers and should not be intended for marketing.

The 2010 regulatory mandate also involved an online registration process for telemarketers starting 15th January 2011 by paying 1000 INR (approx. 13.5 USD), and 9000 INR (approx. 120 USD) as customer education fee. Once validated, the telemarketers are immediately assigned a telemarketer ID, and issued a registration certificate for beginning operations~\cite{traitelemarketerregistration}. Malicious telemarketers found ways to register small fraudulent entities and paid the low registration cost of 10000 INR (approx. 133.5 USD) to access and obtain numbers from the national registry. The malicious telemarketers then use personal phone numbers to send promotional SMS content mimicking the standard peer to peer (P2P) messages which are treated differently than business messages (Note that business messages or bulk ads are 8 times more expensive than P2P SMS). These operations are extremely hard to detect without privacy invasive checks on all customer communications and the telecom operators therefore resort to reactive approaches such as investigation procedures when affected subscribers file complaints with the operator or with the regulatory body. These telemarketers are classified as unregistered telemarketers. In response to increasing cases of fraud and telemarketing from unregistered telemarketers, TRAI capped the number of SMSs that can be sent in a day to 100 per day and limited to 3000 per month for mobile subscribers which was later amended to 200 per day. In the 10\textsuperscript{th} amendment of the regulation, TRAI mandated the usage of \textit{signature solutions} by telecom operators to detect bulk messages having similar string or their variants, specifically those without any specific information of the subscriber. Additional improvements in the amendment included guidelines for block-listing or throttling specific numbers, introducing higher tariffs for bulk messaging, and a web based complaint registration system. In the 13\textsuperscript{th} amendment, the regulation introduced financial disincentives to telecom operators and telemarketers for complaints registered about violations of users' preferences and increased the security deposit needed to be made by the telemarketers to be registered. A recent summary by Tu \textit{et al.} identifies the various techniques currently in use to curb UCC due to robocalls in the United States and similar systems implemented and adopted across the world by various telecom operators~\cite{sokrobocalls}.

Despite incessant measures taken to mandate the use of the NDND registry and regulating the usage of specific number series and SMS header, customer dis-satisfaction continued, indicating the increasing challenges faced by the ecosystem, hinting towards the possible need for revising the regulatory framework.

\subsection{Regulatory Processes and Authors Engagement}
\label{ssec:regulatory_processes}

The main responsibility of the regulators at Telecom Regulatory Authority of India (TRAI) is to help market function. The regulatory change process starts with the initiation of an open consultation to address an identified problem in areas over which the regulators have jurisdiction. The problems are identified through stakeholder feedback, audits, complaints, and active efforts at problem identification. The process starts with a clear presentation of the problem and defines specific issues on which the authority democratically  invites comments from the public.

The responses could be made by stakeholders mentioned in \S\ref{ssec:stakeholders} including active engagement from organizations providing services to the stakeholders, and individual citizens. The responses obtained are openly published in the first stage and an opportunity is given for anyone to submit counter comments. Thereafter, the authority almost invariably chooses to hold one or more open house discussions to hear all points of view in person. The inputs received are summarised, and considered while deciding upon the nature of intervention. This may be a regulation, a recommendation to the government, a direction to the licensees, or a combination of these measures~\cite{direction-trai-2017, direction-trai-2019, direction-trai-2020}. The regulations are accompanied by an explanatory memorandum that sets out the reasoning behind the rules.

The authors in this work engaged with TRAI's initial call for consultation to evaluate the potential feasibility, need, and analyzing the impact of blockchain technologies in the formulated problem setting of spam. These collaborations continued in the public process where inputs were obtained from various stakeholders during the multiple rounds of public comments (and counter comments), and open house events, with the pilot solutions and their feasibility  demonstrated publicly~\cite{comments-coai-2018, comments-bif-2018, comments-itu-apt-2018, comments-cpa-2018, comments-fcso-2018, comments-koan-2018, comments-cerc-2018, comments-ica-2018, comments-microsoft-2018, comments-timesgroup-2018, comments-vfdm-2018, comments-syniverse-2018, comments-cis-2018, comments-pcl-2018, comments-r3-2018, comments-bhavesh-2018, comments-troy-2018, comments-vodafone-2018, comments-reljio-2018, comments-bsnl-2018, comments-idea-2018, comments-airtel-2018, comments-quadrant-2018}. The draft regulation for TCCCPR 2018 was published and invited for a second round of consultation after which some changes were made based on stakeholder input resulting in the updated version of the regulation and an explanatory memorandum was issued addressing the concerns of the stakeholders.

\subsection{Current Workflows}
\label{ssec:business_workflows}

The telecom ecosystem comprises of back and forth communication between the various stakeholders. Principal entities (e.g. online shopping platforms, ride hailing services, restaurants etc..,) partner with a telemarketer with whose help they create a \textit{promotional campaign}. The telemarketer is responsible for the registration of a sender ID for the principal entity which is used in the header format that is displayed to the recipient on their mobile device. Sender IDs allow principal entities to set a recognizable name which helps recipients recognize the brand instantly. Sender IDs/Headers which are equivalent to the domain names in the SMS world need to be registered by the partnering telemarketer who would be running a promotional campaign on the principal entity's behalf. TRAI regulations mandate that the SMS headers for business entities in the countries be of 9 alpha numeric characters of the format \textit{AB-XXXXXX} where \textit{A} corresponds to the letter assigned to the telecom operator, \textit{B} corresponds to the telecom operators' regional circle followed by a \textit{- (hyphen)}. It is mandatory to register a 6 character alpha numeric SMS header. While similar to Internet domain-name registrations, the current process for header registration involves verification of the documentation of the business entity and registering a unique header ID for the same. These registrations need to be shared across all telecom operators and resulted in operational latencies of a few days before the registered header was assigned and made usable to the business entity. 

\begin{figure}
    \centering
    \includegraphics[width=\linewidth]{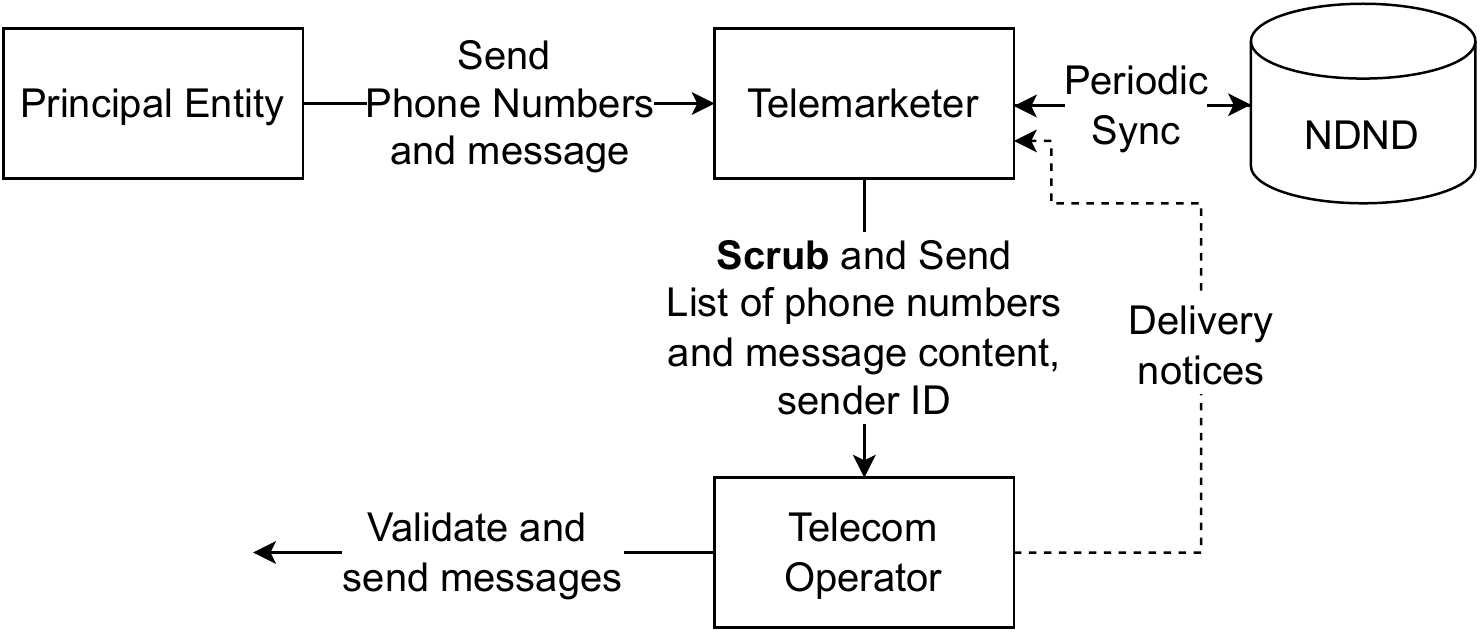}
    \caption{Current Interactions between Principal Entities, Telemarketers, and Telecom operators}
    \label{fig:current_interactions}
    \Description{Figure showing a workflow of interactions between Principal Entities, Telemarketers, NDND database, and Telecom operators showing the scrubbing and validation process being done by the telemarketer before sending the messages and phone numbers to the telecom operator.}
\end{figure}


Today, The principal entity provides the telemarketer with a list of phone numbers who are the target of the promotion either as a file or via an API provided by the telemarketer. The telemarketer then removes the numbers registered with the DND service from the original list in a process known as \textit{\textbf{scrubbing}} and proceeds to send the final list and promotional message to the a partner telecom operator to be sent under a specific sender ID or a generic sender ID for promotional SMSs registered to the telemarketer. The telecom operator who receives the task, known as the Originating Service Provider (OSP), further splits the list provided into regional circles based on the registration of the phone numbers and delegates the task of sending the SMS to the telecom operator at a given circle, known as the Terminating Service Provider (TSP), who then sends the required promotional SMS. The TSPs optionally scrub the phone numbers again before sending the SMS. We present these interactions in Figure~\ref{fig:current_interactions}.

To register or update preferences, a subscriber logs into a centralized web based portal provided by TRAI or a portal provided by their respective telecom operators. In the current system, the preferences are sent daily as an update from each telecom operator to the national do not disturb registry. The registry updates are then periodically synced by the respective telecom operators and the telemarketers. The existing system takes more than 24 hours to register the preferences of the user and requires up to seven days to enforce the revised preferences at the TSPs~\cite{traidec2018}.

A complaint can be registered for an unsolicited commercial communication by a mobile subscriber using the mobile apps provided by TRAI or via their respective telecom operators' web, IVR or SMS based systems. Once a complaint is registered by the subscriber, the TSP verifies the registration status of the complainant in the National DND registry and compares the information provided using semi-automated procedures in the logs to identify the OSP, and telemarketer who have initiated the process. The parties involved along with TRAI verify the complaint and present required proofs of SMS being sent and the content in a specific promotion. On confirmation of UCC, the telecom operator and the telemarketer are levied fines. The total turn around time for a complaint to be addressed in the current process is close to 7 working days and involves heavy coordination between the telemarketer, and the telecom service providers (TSPs, OSPs) involved in SMS message flow. The problem further intensifies due to customer data privacy and can only be addressed by nodal officers of the respective telecom operators from each circle.

\subsection{Challenges with Current Workflows}
\label{ssec:challenges_current_workflow}

The problem of UCC has continuously been a challenge for regulators, mobile subscribers and telecom operators alike. The National DND registry which consists of all the phone numbers of the customers and their corresponding preferences is stored in plain text and can be accessed by all registered telemarketers who create periodic local copies for scrubbing purposes. TRAI amended the UCC regulations in India and moved from a binary mode of setting customer preferences i.e. \textit{full block of promotional content}, and \textit{no restrictions on promotional content}, to enabling \textit{partial blocks}. Using the partial block features, the subscriber can partially block content from seven predefined set of categories which are as follows: 1) Banking, insurance \& financial products, 2) Real estate, 3) Education, 4) Health, 5) Consumer goods \& automobiles, 6) Communication, broadcasting \& entertainment \& 7) Tourism. The introduction of this resulted in the need for telemarketers to classify the principal entities under one or more of the respective categories. Mobile subscribers blocking a specific category due to the UCC generated by a specific principal entity also end up blocking principal entities who they would have otherwise been interested in. This opens up a need for nesting the categories to multiple levels of sub-categorization.

A major problem for unsolicited commercial communication is due to possible collusion between malicious and unregistered telemarketers. The availability of all phone numbers for a small registration fee and security deposit to TRAI allows malicious telemarketers to leak phone numbers and collude with unregistered telemarketers or entities to violate the guidelines set of telemarketing and send out SMS or make calls from phone numbers which are registered for personal use. Therefore there is a need for storing the numbers in the DND registry in a privacy preserving format which the telemarketers can then scrub their phone number lists across the DND service protecting user privacy, instead of periodically downloading the data present in the NDND registry for scrubbing.

The current registration of the SMS headers/sender IDs for the principal entities opens up tremendous opportunities to \textit{spoof} customers with sender IDs which sound and look very similar. For example, a leading bank in India, the State Bank of India (SBI) might be registered as \textit{STABAN}. However, another malicious entity under the name of \textit{SBI Banquets} can register themselves as \textit{SBIBAN} and send out fraudulent SMSs posing as the bank. Additionally, many principal entities might not have a 6 character name for e.g. \textit{Ola, Uber} (popular ride sharing services) which makes telemarketers register them with additional characters such as \textit{OLACAB, UBERCA}. Similar issues exist with principal entities whose names are much longer and cannot be abbreviated to six characters. Many of the large organizations in India have multiple businesses in different sectors which makes it hard to establish a brand identity with a flexibility of six characters. Additionally business entities use multiple telemarketers and SMS gateway APIs as a fallback in case of failure thereby making it hard for the user to associate the message with the brand. For example, SMSs from the bank stated above might be received on the subscribers' mobile phone as \textit{VM-STABAN}, \textit{AD-STABAN} etc.., which appear under different SMS inboxes for the mobile subscriber making it difficult to keep track of all the SMSs from a specific principal entity in a common place. These challenges provide us with an opportunity to implement national header registration features similar to domain name systems and their registrations for the web~\cite{mockapetris1988development}. Being able to track and verify the message flows between principal entities, telemarketers, and telecom partners, it becomes possible to enable mobile subscribers to efficiently manage their SMS inboxes through improved flexibility in display of the sender ID.


The current processes for a subscriber to register their preferences and for them to come into effect can take up to seven working days due to synchronization delays. Similarly, a registered complaint can take up to 7 days to be resolved and needs to be resolved manually in most cases with extensive collaboration between the different telemarketers, telecom operators, and subscribers. This opens up the need to simplify the operations and automate the way in which data is shared and used by the respective stakeholders. During complaint resolution it also becomes very difficult for the principal entities and telemarketers to prove acquisition of consent. In many cases these proofs are presented as paper documents, screenshots with insufficient information or database entries made across a subscribers' record with no way to verify the validity of the claim. This brings in a need for a verifiable and digital way of acquiring, storing and using consent from the customers while giving the customers an accessible way to revoke it.

Since the introduction of the complaint registration procedure over 2 million complaints have been registered by consumers with an average of 40000 per month. As a result, telecom services to more than 1.4 million phone numbers have been disconnected and 460000 numbers have been blocked so far. Despite so many disconnections, levy of huge fines and blocklists of unregistered telemarketers \& violators, the UCC problem is still not fully under control. Motivated by these challenges (summarized in Table~\ref{table:challenges_and_design}), we propose a blockchain based solution aimed at addressing various privacy, security, and operational challenges.

\begin{table*}[t]
\small
\caption{Challenges faced by different stakeholders grouped by design goals and deployment status. \\ (\CIRCLE: Completely deployed, \LEFTcircle: Partial deployment)}
\resizebox{\linewidth}{!}{
\begin{tabular}{|l|l|l|l|}
\hline
\textbf{\#} & \textbf{Stakeholder} & \textbf{Challenges} & \textbf{Addressed by} \\ \hline
\multicolumn{4}{|l|}{\cellcolor{gray!25}\textbf{Real time}} \\ \hline
R1 & Subscribers & Lengthy time for preference registration to take effect. (7 days) & \CIRCLE~ [\S\ref{sssec:blockchain_network}, \S\ref{sssec:smart_contracts}] \\ \hline
R2 & Subscribers & Lengthy time for complaint resolution. (7-14 days) & \LEFTcircle~ [\S\ref{sssec:blockchain_network}, \S\ref{sssec:smart_contracts}] \\ \hline
R3 & Telemarketers & Ensuring the usage of the latest data for scrubbing phone numbers & \CIRCLE~ [\S\ref{sssec:blockchain_network}, \S\ref{sssec:scrubbing_nodes}, \S\ref{sssec:smart_contracts}] \\ \hline
R4 & Operators \& Telemarketers & Quick update and maintenance of user preferences & \CIRCLE~ [\S\ref{sssec:blockchain_network}, \S\ref{sssec:smart_contracts}] \\ \hline
R5 & Operators & Coordination with telemarketers & \CIRCLE~[\S\ref{sssec:blockchain_network}, \S\ref{sssec:scrubbing_nodes}, \S\ref{sssec:smart_contracts}] \\ \hline
\multicolumn{4}{|l|}{\cellcolor{gray!25}\textbf{Flexibility}} \\ \hline
F1 & Subscribers & Unavailability of fine grained control of preferences & \LEFTcircle~ [\S\ref{sssec:preference_registry}] \\ \hline
F2 & Principal entities & Registration of longer/shorter SMS Header/Sender ID & \LEFTcircle~ [\S\ref{sssec:blockchain_network},  \S\ref{sssec:smart_contracts}] \\ \hline
F3 & Principal entities & Brand identification irrespective of mediating links & \CIRCLE~ [\S\ref{sssec:blockchain_network}, \S\ref{sssec:header_content_registry}, \S\ref{sssec:consent_registration}] \\ \hline
\multicolumn{4}{|l|}{\cellcolor{gray!25}\textbf{Security \& Privacy}} \\ \hline
S1 & Telemarketers & Avoiding usage of plain text phone number sharing & \CIRCLE~ [\S\ref{sssec:ndnd_privacy}] \\ \hline
S2 & Operators & Ensuring data privacy and protecting consumer data & \CIRCLE~ [\S\ref{sssec:ndnd_privacy}, \S\ref{sssec:scrubbing_nodes}] \\ \hline
S3 & Subscribers & Prevent access of phone number data by unregistered telemarketers & \LEFTcircle~ [\S\ref{sssec:ndnd_privacy}] \\ \hline
S4 & Operators & Prevent unregistered telemarketers from sending promotional SMS/calls & \LEFTcircle~ [\S\ref{sssec:end_user_applications}] \\ \hline
S5 & Principal entities & Protection from spoofing attacks due to similar sounding sender IDs & \CIRCLE [\S\ref{sssec:blockchain_network},  \S\ref{sssec:smart_contracts}] \\ \hline
\multicolumn{4}{|l|}{\cellcolor{gray!25}\textbf{Automation \& Usability}} \\ \hline
A1 & Operators & Validating complaint registrations \& root cause analysis & \LEFTcircle~ [\S\ref{sssec:blockchain_network}, \S\ref{sssec:header_content_registry},  \S\ref{sssec:smart_contracts}] \\ \hline
A2 & Subscribers & Avoiding inbox clutter due to multiple sender IDs & \LEFTcircle~ [\S\ref{sssec:end_user_applications}] \\ \hline
A3 & Auditors \& Regulators & Consent request \& acquisition is unverifiable & \LEFTcircle~ [\S\ref{sssec:blockchain_network}, \S\ref{sssec:header_content_registry}, \S\ref{sssec:consent_registration}, \S\ref{sssec:smart_contracts}] \\ \hline

\end{tabular}
}
\label{table:challenges_and_design}
\end{table*}

\subsection{Design Goals}
\label{ssec:design_goals}

With the understanding of the current workflows in the telecom ecosystem described in \S\ref{ssec:business_workflows} and the various challenges they pose described in \S\ref{ssec:challenges_current_workflow}, we outline below the design goals, summarized by the grouped categories shown in Table~\ref{table:challenges_and_design}. We set out to achieve our goals with the proposed solution in \S\ref{sec:proposed_solution}.

\begin{enumerate}
    \item \textbf{Real time or Near Real time operations}: Subscribers, and principal entities both face significant challenges due to the long operational times involved in critical operations. Subscribers face lengthy times for the registration of their preferences (1-7 days), and for them to take effect. This is due to synchronization challenges between the telecom operators, and the TRAI operated NDND preference registry. Principal entities face similar challenges with the registration of headers to send promotional messages. We aim to reduce these multi-day latencies to real time, or near-real time latencies when operating at scale.
    
    \item \textbf{Improving Security and Privacy}: Telemarketers and telecom operators periodically obtain the plain text information of subscribers in the NDND registry and their preferences for promotions. Malicious telemarketers obtain the information of these subscribers and use personal devices to perform  telemarketing operations. We aim to prevent the usage, and sharing of plaintext phone numbers, and prevent risks from malicious attackers compromising and leaking the subscriber phone numbers, and preferences in the NDND registry.
    
    \item \textbf{Improved Compliance and Flexibility}: Subscribers register complaints for UCC through various means such as through the official mobile app managed by TRAI~\cite{iostrai, androidtrai}, with the telecom operator providing them the services through a web portal, IVR, customer complaint ticket. The registered complaints are usually addressed within 7-14 working days. We aim to reduce these latencies, improve the compliance of telemarketers to adhere to subscriber preferences. Successful intervention of the proposed system would result in a reduced number of complaints from registered telemarketers. We aim to design the solution to be easily extensible allowing future fine grained preferences for customers, improved header registrations, and protect brand reputation through fraud prevention. 
\end{enumerate}

\section{Proposed Solution - Telechain}
\label{sec:proposed_solution}

The most common approach today in India, and other countries, to the problem of UCC, is to investigate spam after a violation is detected or a complaint is filed by the subscriber. In contrast, our solution helps establish ex-ante compliance for each message or call sent to the mobile subscriber -- thus effectively ensuring that all the promotional messages sent to a particular user are consistent with preferences and recorded consent of that user.

Our choice of a consortium blockchain in the design of Telechain is motivated by our goals described in \S\ref{ssec:design_goals} to have secure and near real time exchange of data between the various stakeholders. In addition, the pro-active approach to policy enforcement and transparency based on collective verifiability and smart contracts enabled by blockchain makes it a compelling design choice. 

\subsection{Components of the proposed system}
\label{ssec:components}

\subsubsection{\textbf{Blockchain network}}
\label{sssec:blockchain_network}

A consortium blockchain network is created between the participating entities i.e. telemarketers contributing physical  nodes in the blockchain, telecom operators, third party service providers and a regulatory authority. Any update of the customers' preferences in the NDND registry via the mobile app, web portal or via telecom operators customer care systems (described in \S\ref{sssec:end_user_applications}) is encoded as a transaction in the system. Each one of the participants in the network maintains a copy of the blockchain state including the DND register which gets updated in the servers of the participating entities thus making each update of preferences by the subscribers near-real time and reduces the time from almost seven days to a matter of minutes. The usage of blockchains and establishing the network addresses the current latency challenges and issues due of data staleness (R1-R5), improves policy adherence by ensuring strong accountability and transparency addressing the S5, A1, A3 challenges, facilitates the necessary registrations of headers, templates, and consent (F2, F3) presented in Table~\ref{table:challenges_and_design}.

\subsubsection{\textbf{Improving privacy of NDND Registry}}
\label{sssec:ndnd_privacy}

To improve privacy and security, subscriber information shared on the chain is always hashed and cryptographically signed by the respective nodes originating the update. The use of blockchain changes the maintenance of the DND registry from a centralized setting to a decentralized and replicated setting. While standard access control measures prevent leakage of information to unauthorized telemarketers, the existence of hashed information  prevents a malicious telemarketer from leaking actual phone numbers to an unregistered telemarketer who carries out advertising through personal cellular devices or a telecom operator poaching subscribers addressing the challenges S1, S2, and S3, presented in Table~\ref{table:challenges_and_design}. Each number present in the DND registry is mapped to the name of the telecom operator and relevant metadata.

While this approach allows large telemarketers to make their operations more efficient in sending promotional SMS, it makes it harder for smaller telemarketers who cannot afford to dedicate a server on their behalf to participate in the network. Previously, smaller telemarketers filtered the list of customers from the do not disturb registry by downloading the entire data of the registry and possibly manually checking the customer preferences for each promotional campaign. However, this approach would no longer work with the new solution since the hashed information provided to the telemarketers is not usable with their current operations. To resolve this problem, we introduce \textit{scrubbing nodes} on the blockchain network providing scrubbing services for a minimal fee. Larger telemarketers, telecom operators and infrastructure operators provide scrubbing services to smaller telemarketers.

\subsubsection{\textbf{Scrubbing Nodes}}
\label{sssec:scrubbing_nodes}

The scrubbing nodes have read only privileges to the blockchain network and are responsible for providing scrubbing services for a small fee. The result of the scrubbing would be a token that's furnished to the telemarketer and the proposal of a transaction to the blockchain while following a specification that's agreed upon by the participants in the network. The telemarketer can submit this token along with the promotional SMS, registered header, and template IDs, directly to the telecom operator who can retrieve the scrubbed file and send the promotional SMSs to the list of valid subscribers. The design decision to return a token instead of the scrubbed result $S = \mathbb{L} - \mathbb{C}$, where $\mathbb{L}$ corresponds to the set of subscriber identities and $\mathbb{C}$ corresponds to the set of subscribers in the NDND registry, is intentional to prevent probing attacks where telemarketers aim to identify subscriber preferences by sending one request at a time. The scrubbed result at its simplest binary preference level is represented by the set difference $S$. The scrubbing nodes coupled with the blockchain network ensure the recency of the data being operated upon, easy interoperability with existing telecom operators, prevent unauthorized access to the subscribers' consent and preferences and protects their privacy addressing the challenges R3, R5, and S2, presented in Table~\ref{table:challenges_and_design}. 


\subsubsection{\textbf{Header \& Content Template Registry}}
\label{sssec:header_content_registry}

The blockchain network handles the registration of headers (sender IDs) and maintains the ownership mapping between the principal entities and the headers registered by the entity. The telemarketer operating the promotional campaign on behalf of a principal entity is required to register the content of the message(s) used as a part of the campaign. The content registration is also allowed on the system as \textit{templates} which are message strings that have placeholders and can be replaced by specific information related to the subscriber when the SMS is sent. Here is a sample content template.

\textcolor{gray}{\say{\textit{Dear <\%..\%> Head to the nearest <\%..\%> store and avail a flat <\%..\%>\% off using coupon code <\%..\%>.}}}

The registration of headers and templates by the principal entities however need manual verification by the telecom operators for business compliance. Once verified and registered into the blockchain, it provides principal entities ownership of their brand header and verification to operators and telemarketers after the correct scrubbing process about the correctness of the message delivery addressing F3, enabling easier complaint tracking (A1), and establishing auditability of the actions (A3).

\subsubsection{\textbf{Preference Registry}}
\label{sssec:preference_registry}

A challenge in the existing system was the long duration needed for a customer 
preference to take effect. With the introduction of the blockchain, a mobile subscriber can login via their respective telecom operators self-help portals or the mobile apps to update their preferences and either block/unblock specific promotional categories and sub-categories. The changes are proposed as a transaction by the telecom operator, or TRAI enabling the dashboards after successful user authentication. The transaction on successful validation is used to update the database records of each of the participants with the new preference information set by the customer effectively reducing the time needed for the preferences to be in effect \textit{improve from seven days to a few minutes}. Additionally, the flexibility of the scrubbing services and its integration into the blockchain enables the preference registry to support fine grained preferences addressing the challenge F1, and enabling fine grained preference control.

\subsubsection{\textbf{Consent Registration \& Acquisition}}
\label{sssec:consent_registration}

To overcome the problem of unsolicited commercial communication by telemarketers to potential customers, it becomes highly important to have an electronic registration of consent of the customer. This process is made in two independent phases, the first phase involves the registration of a \textit{consent template} by the principal entities which maps it to their registered headers. The registered message will contain the link to privacy policy, terms and conditions, message frequency and request the customer for approval before being contacted for any promotional purposes. The second phase is the \textit{consent acquisition} phase where the registered consent template is sent over SMS to the mobile subscriber along with a one time password or a custom web link. The subscriber can read the conditions and choose to either provide the principal entity with the OTP or click on the link and go through the necessary steps to provide consent to the principal entity. This provides a digital trail of actions (addressing A3) and SMS interactions that resulted in the verifiable informed consent being acquired by the principal entity (addressing F3) for their customers.

\textcolor{gray}{\say{\textit{Thank you for shopping with <\%..\%>. Please click on the following <\%link\%> if you'd like to receive updates regarding offers \& promotions. Terms \& privacy can be found here. You may receive up to 4 sms/month. Your consent OTP: <\%..\%>}}} - sample consent template.

While consent reflects on an explicit preference, the exercise of preferences does not imply consent. For example, a subscriber consenting to receive promotional offers from a restaurant is only restricted to that specific restaurant and not to the overall preference for receiving promotional offers from \textit{any} restaurant.

\subsubsection{\textbf{Smart contracts \& agreements}}
\label{sssec:smart_contracts}

Smart contracts are programmatic expressions of business workflow conditions which help in pro-active policy enforcement without the need for a third party. The developed contracts are audited by the participants of the network and across a correctness test suite which implements the various conditions described above such as the \textit{header, content \& consent registrations}, \textit{scrubbing requests} and are validated by the participants of the network. These are programmatic encoding of business logic which are critical and facilitate the operations on the blockchain network. 

\subsubsection{\textbf{End user applications}}
\label{sssec:end_user_applications}

Each of the participating nodes (telecom operators, telemarketers, scrubbing nodes) in the blockchain network expose interfaces for their existing business processing systems to interact with the blockchain infrastructure. These include web portals, mobile apps, Unstructured Supplementary Service Data (USSD), and SMS services provided by telecom operators where a mobile subscriber can register complaints, update preferences, and in addition can see a detailed list of consents provided by the customer. Similarly, there are supporting applications provided for smaller telemarketers where the promotion campaign can be registered on behalf of a principal entity which allows the telemarketers to obtain a token to be furnished to the telecom operator in addition to also providing them an easy way to request a telecom operator to run the promotional campaign. End user applications such as messaging apps can continue to implement local spam and aggregation filters to avoid inbox clutter addressing A2 while applications like TrueCaller installed on client mobile devices can block unregistered telemarketer SMS and calls.

\begin{figure}[ht]
\includegraphics[width=\linewidth]{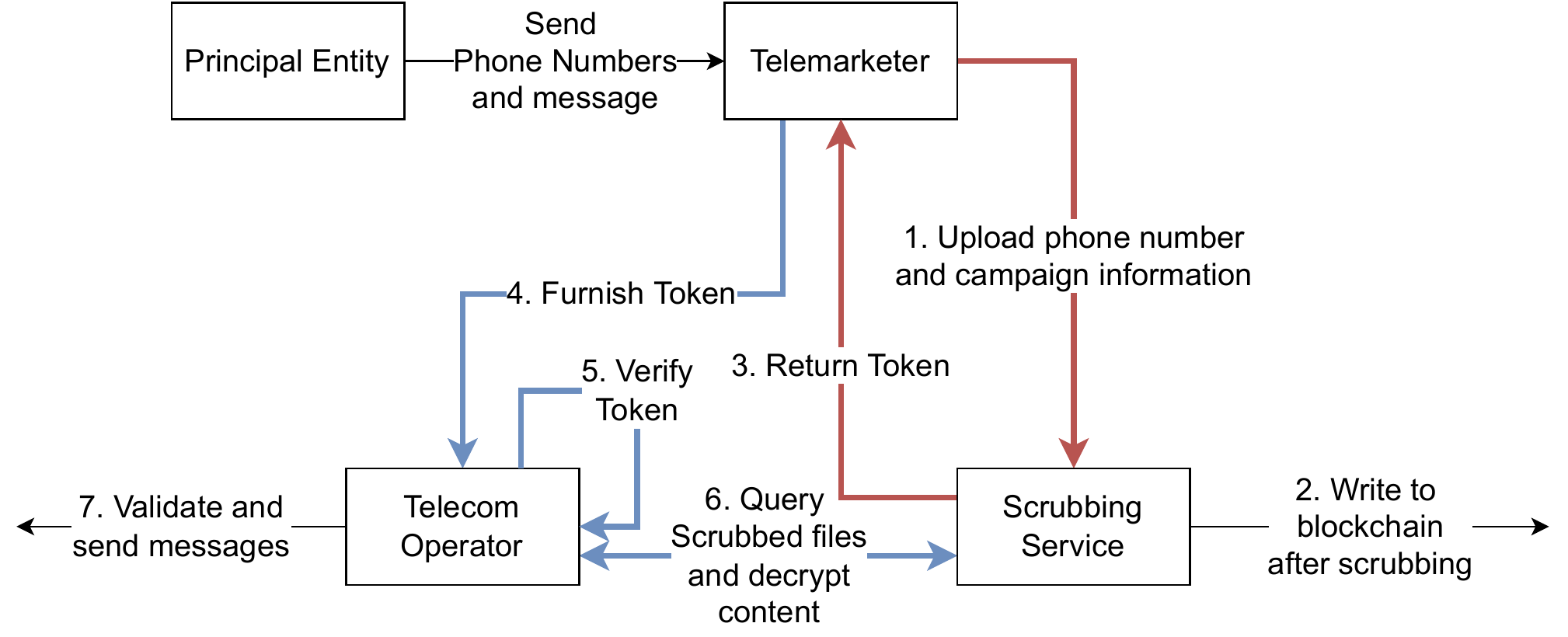}
\caption{Figure showing the new workflow among stakeholders involving generation of a scrubbing token and the usage of scrubbing services. The solid \textcolor{red}{red} lines indicate the scrubbing service workflow. The scrubbing operations can be performed through any registered blockchain participant providing the scrubbing services. The solid \textcolor{blue}{blue} lines indicate the post-scrubbing workflow which is modified from the workflow presented in Figure~\ref{fig:current_interactions} for final message delivery.}
\label{fig:new_workflow}
\Description{Figure showing a new workflow among the stakeholders of the telecom ecosystem, the usage of the updated scrubbing process and the integration of the blockchain network for interoperability preventing the need for plaintext communication between the stakeholders.}
\end{figure}

\subsection{Decentralized Workflows}

Keeping in mind the business workflows detailed in \S\ref{ssec:business_workflows}, and the desire to create minimal disruptions to existing business processes, we design a new decentralized workflow leveraging the components presented in \S\ref{ssec:components} to tackle the challenges summarized in Table ~\ref{table:challenges_and_design} and \S\ref{ssec:challenges_current_workflow}. The registration / on-boarding of each of the components involved in the workflows are as follows:

\subsubsection{\textbf{Telemarketer Registration}}
\label{sssec:telemarketer_registration}

Telemarketers register with TRAI using the existing application process, pay the required one time fee and a security deposit after which they receive a telemarketer ID. The telemarketer ID along with the payment information is stored in a database provided by TRAI to which the \textit{read-only} credentials are shared over the network. Telemarketers willing to participate in the blockchain network can create a server and request to connect to the blockchain network by furnishing their a self signed cryptographic identity along with the telemarketer ID and payment receipt numbers as a transaction. The information corresponding to the telemarketer ID is verified by each of the participants according to the policy of the network. On successfully validating the information on the network and ascertaining the identity and signature, the server node is added to the network as a participant in the blockchain. The telemarketer node then reconstructs the blockchain by fetching information from the peers in the network and validates them before finally reaching a state consistent \textit{global state} matching the other participants in the network after which the added node can participate in the network by issuing new transactions.

Telemarketers who do not participate in the network as direct participants delegate their registration to the network via any third party services, detailed later in the paper, which create a local authentication system for the telemarketer, maintain their cryptographic credentials and register them by creating a transaction in the network.

\subsubsection{\textbf{Principal Entity, Header \& Content Registration}}
\label{sssec:principal_header_content_registration}

Principal entities are registered along with the \textit{header} registration processes via \textit{third party services} or telemarketers offering header \& content registration services.  Similar to domain name registration, a principal entity registers the headers of their choice which appear as sender ID to the mobile subscribers. The registration of the principal entities is done using their business name while mapping it to the chosen header(s) as a transaction on the blockchain. These entities are granted access to their information using the authentication on the third party services.

The principal entities after confirmation of their header registrations delegate/configure the ownership or access of their registered headers to any of the registered telemarketers by using the telemarketer ID or searching for the same via the web interface provided by the third party service for the same. This allows the partner registered telemarketers to create campaigns on behalf of the principal entity. The access delegation of a header to the telemarketers by the principal entities is recorded as a transaction over the blockchain. Once delegated, the telemarketers can register promotional content templates or the complete content of the promotion against the header of the principal entity by signing the one way hash of the templated content and the header used as the message. A transaction is issued on the blockchain by the principal entity with ownership of a specific header proxied through a blockchain participant offering content registration services. A \textit{content token} is provided to the telemarketer which is furnished to the partnering telecom operator along with the template and the \textit{scrubbing token} for message delivery as shown in Figure~\ref{fig:new_workflow}.

\subsubsection{\textbf{Customer \& Consent Acquisition}}
\label{sssec:customer_consent_acquisition}

During customer acquisition and storing the contact information of the customer (for eg. at the point of sale counter), principal entities can use the APIs provided by their partner  telemarketer or the services offered by third party consent registration and maintenance services to trigger a one time password (OTP) along with mandatory information as described previously in \S\ref{sssec:consent_registration}. The trigger of the consent acquisition SMS based on a registered consent template as described in \S\ref{sssec:principal_header_content_registration} is recorded as a transaction on the blockchain. Similarly, the grant of the consent either based on OTP or a provided web link to the customer is recorded on the blockchain as a transaction which on successfully being validated adds the consent mapped to the header and the customer for future promotional campaigns by the principal entity.

\subsubsection{\textbf{Third Party Services}}
\label{ssec:third_party_services}

Third party services are value added services offered by nodes participating in the network which can offer any of the following services. The services include 1) registration of telemarketers, principal entities and headers, content and consent,  2) authentication to third party services, scrubbing as a service using preferences and consent for principal entities, and 3) execution of promotional campaigns and forwarding the requests to the telecom operators. The third party service providers create and register identities for the telemarketers and provide them an authentication system for further usage of the system. These nodes behave as proxy nodes for their users to interact with the blockchain and provides the necessary web applications or API services.

\textbf{Scrubbing} is a mandatory business process for sending out any promotional SMSs with failures to comply with the mandate resulting in heavy penalties including ban of operations. The scrubbing process can be performed by the  telemarketer nodes participating in the blockchain on their respective server machines. Telemarketers not actively participating in the network can use third party scrubbing services which are participants on the network. The scrubbing operation takes in an input list of phone numbers, the telemarketer ID, content template ID and the header of the principal entity. After validating the ownership of the telemarketer ID with the header ID, and the content ID with the header ID, the service creates a one way hash of each number and validates the preferences and the consents listed across the entry of the particular phone number from the current global state of the blockchain. If the user has either consented to receiving promotional content from the principal entity or has the category of promotion as a valid preference, the number of the user is stored in multiple files of valid phone numbers where each file contains the numbers corresponding to a specific telecom operator. Similarly, the rest are stored in corresponding invalid files.

Each one of these files based on the telecom operator are encrypted using the public key of the telecom operator and the corresponding links to the file locations for each telecom operator is provided along with a signed digest of the respective files thereby proving immutability and allowing access to only the intended parties. In addition the transaction includes a signed token and current hash of the global state which is issued to the blockchain network. The participants of the network validate the contents of the transaction by verifying the digests of the encrypted files and corresponding signatures. The telecom operators after validation take the corresponding token issued in the transaction and queue it to the pipeline of promotional campaigns to run. The token and the transaction ID proposed by the scrubbing node is presented to the telemarketer who requests the telecom operator(s) to begin the campaign by furnishing the token and signing it with their cryptographic identity. The request to initiate the campaign is recorded on the blockchain and the promotional campaign takes effect with each telecom operator changing the status of the promotional campaign to \textit{completed}.

The business workflows post registration of the telemarketers, principal entities are illustrated in Figure~\ref{fig:new_workflow} where the process starts with the principal entity partnering with a telemarketer. Similar to the current process, the principal entity sends a list of phone numbers and the message content, and a registered template ID to the telemarketer who scrubs the data across a third party scrubbing service or on their own while initiating a corresponding transaction on the blockchain. The generated token from the scrubbing process is furnished to the telecom operator directly or via third party services. The telecom operator validates the token, reads the corresponding list of phone numbers to send the promotional SMS and validates the transaction status by optionally re-scrubbing them locally before sending out the promotional SMSs to the mobile subscribers in their network. The telecom operator(s) involved in the message delivery then eventually attach a delivery report to the telemarketer initiating the campaign indicating the total number of successful messages delivered which can be used for subsequent billing purposes.

\subsection{Challenges addressed}
\label{ssec:challenges_addressed}

With the implementation of the proposed system using the components introduced in \S\ref{ssec:components}, we addresses many of the challenges posed in Table~\ref{table:challenges_and_design}.

By hashing the information available in the centralized NDND registry~\cite{dash2009national}, replicating, and decentralizing it using blockchain, we ensure that the preference registration and any  \textit{updates are instantly in effect across all the participants in the network} removing the periodic synchronization in favor of distributed consensus protocols which are critical to blockchains~\cite{castro1999practical, sukhwani2017performance}. This reduces the complex and time consuming process in the current approach where the data is periodically downloaded and synchronized by various stakeholders. The proposed system and format of hashed data storage \textit{protects sensitive and personally identifiable consumer data} (phone number) from being leaked. Malicious telemarketers and colluding unregistered telemarketers can no longer make sense of the downloaded hashed format.

The usage of the blockchain is ideal for the given problem as it \textit{automates the coordination between various stakeholders} and their corresponding cryptographic identities on the system. The mandatory need for scrubbing ensures that the telemarketers do not ever know a \textit{filtered} list of phone numbers which could help malicious entities collude with unregistered telemarketers. The segregation of the files by the scrubbing operations into respective files for telecom operators \textit{greatly simplify the operations} and the need for the partner telecom operator to share data with other telecom operators for executing the promotional campaign. Throughout the entire process, the phone numbers and other sensitive information is shared in a hashed manner thereby \textit{protecting the privacy of the  subscribers}. 

The availability of all the registered entities, their categories of operation and recursive sub-categories are now available to each of the nodes participating in the network. This information enables scrubbing service providers and preference registry systems to give users \textit{much more fine grained control of their preferences} than the coarse categorisation into seven categories. It also allows any mobile subscriber to clearly view, manage and revoke the consents granted by them to various entities from a single location. Principal entities because of the new header registration process and the removal of the six character limit can now clearly establish their brand identities. Apps installed by mobile subscribers on their devices now can organize their message inboxes by the name of the business entity making it cleaner and easier to look for SMSs and avoiding clutter. The digital trace and availability of the consent makes it easier for auditors to identify and penalize violations based on complaints registered by the user.






\section{Implementation}
\label{sec:implementation}

\subsection{Implementation of the Blockchain Network}
\label{ssec:implementation_blockchain_network}

\begin{figure}[ht]
\includegraphics[width=\linewidth]{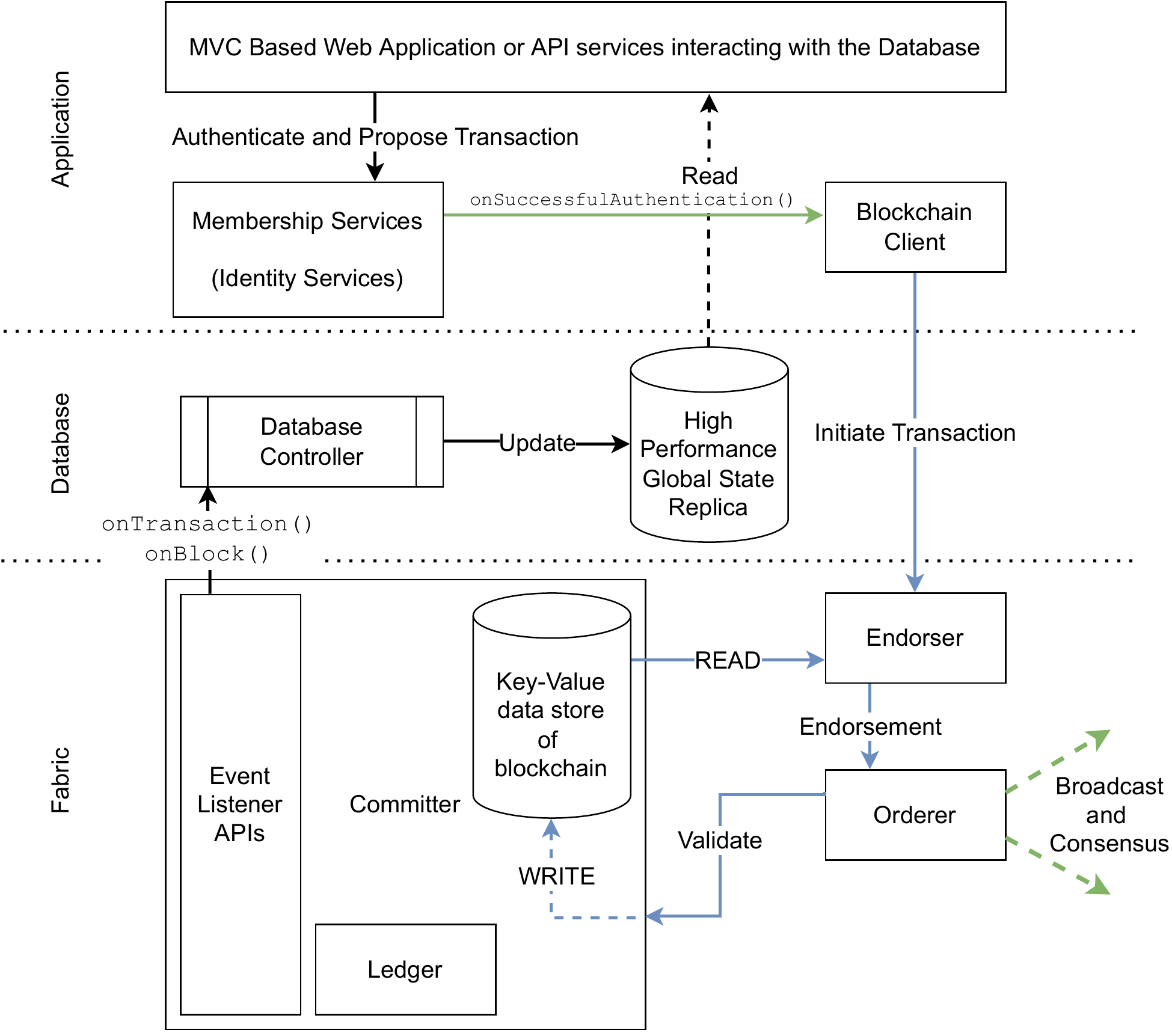}
\caption{Blockchain participant node architecture indicating the web based application layer which includes an authenticated service proxying the requests made by the application users as a blockchain client. The database layer includes event driven approaches to replicate the global state of the blockchain, especially used by scrubbers for high performance tasks, layered over Hyperledger Fabric infrastructure. The blue arrows indicate the sequence of events in the blockchain.}
\label{fig:node_architecture}
\Description{Figure showing the tiered breakdown of the participant node in the Blockchain. The figure shows the application, database, and the underlying Hyperledger Fabric layer with various database and communications operations performed by the node.}
\end{figure}

To test the feasibility of our proposal, a pilot implementation of the consortium blockchain network is built using Hyperledger Fabric, an open source consortium blockchain framework under the Hyperledger umbrella of the Linux Foundation~\cite{Androulaki:2018:HFD:3190508.3190538}. Individual nodes of the telecom operators, a large telemarketer node, scrubbing service node and an observer node for the regulator (TRAI) form the permissioned blockchain network. \textit{Endorsement policies} of the network are policies through which designated system administrators specify conditions for various types of transactions to be endorsed. It is also possible to specify custom endorsement policies, for example, a transaction to register a principal entity and their corresponding headers need to be endorsed by each of the telemarketer nodes on the network, the observer node and \textit{at least} one of the telecom operators. Figure~\ref{fig:node_architecture} shows the modules and implementation architecture at each physical node in the network.

In our implementation a majority of the participants on the network must endorse a transaction initiated by a client. During the endorsement, the designated peers according to the endorsement policy execute the transactions and share the output as the endorsement. The result of the endorsement is a cryptographically signed \textit{read-write set} which includes the sequence of operations to perform on an underlying key-value  database (CouchDB/LevelDB) to commit the transaction. This is indicated as the \textit{Endorser} in Figure~\ref{fig:node_architecture}.

The read-write sets are then ordered by an ordering service which is a part of the Fabric framework. The consensus protocol which runs between the orderers of the network order the endorsed transactions and group them together into a \textit{block} which is broadcast to all the participants of the network who then deterministically validate the endorsements \& the read-write sets. On successful validation a block is appended to the local ledger of each of the participants in the network and chained to the hash of the previous block. The corresponding key-value pairs indicated in the read-write sets are modified accordingly to achieve a new global state. This is indicated as the \textit{Committer} in Figure ~\ref{fig:node_architecture}.

\subsection{Integrating Existing Identity Systems}
\label{ssec:existing_identity_integration}

The organizations participating in the network have their own centralized identity systems such as Azure Active Directory, OpenLDAP, Apache directory, IBM Tivoli, or similar. The credentials for each stakeholder in the system is maintained by the respective identity systems. A successful registration of a principal entity or telemarketer via a participant node on the network creates a cryptographic identity using standard public key infrastructure (PKI) methods and grants authentication to the network using digital signatures. Each node participating in the network stores their credentials and publishes the necessary public key information to other identity server nodes establishing a relationship between the identities of different organizations and the corresponding cryptographic keys needed for validation of transactions initiated over the network. This is indicated as the \textit{membership service} in Figure~\ref{fig:node_architecture}.

\subsection{Implementation of the Scrubbing Services}
\label{ssec:implementation_scrubbing_service}

Scrubbing is a mandatory operation that needs to be performed on an input list of promotional phone numbers and other mandatory information as described in \S\ref{ssec:third_party_services}. These services are democratic and can be offered by anyone who willingly participates in the blockchain network after regulatory approval. The regulatory approval for the participants in the network is obtained after successfully scrubbing a standardized suite of test cases put together by the consortium and audit of the software system being used. The scrubbing operation reads the value associated with each key i.e. (hash of the phone number) from the consent and preference registries which are locally maintained in each participating nodes' key value data store as indicated in Figure~\ref{fig:node_architecture}. The scrubbing implementation can connect to the same database and filter the numbers from the input list provided and create the output files and encrypt them using the corresponding public keys of each of the telemarketers which can be obtained from the membership service.

To improve the speed of filtering through this data, we use the \textit{event listeners} provided by the committer to listen to the completion events of block \& transaction confirmation and mirror the updates to fabric's key value stores in a flexible format and re-indexing the data to optimize any queries over an open source in-memory database like \textit{redis} with efficient and fast set difference implementations (\texttt{SDIFF}) compared to performing the same within document stores~\cite{redisbenchmark}. This enables extremely quick scrubbing results and the service is metered. Prior to this system, scrubbing was a time consuming operation as it involved the download of the entire data from NDND registry and reconciling the updates with their local databases before continuing to perform the telemarketing activities. Each scrub performed by the telemarketers was considered valid for a week thereby making it difficult for subscriber preferences to take immediate effect. The new system prevents this and enables immediate effect on subscriber preferences.





\subsection{Integrating Existing End User Applications}
\label{ssec:end_user_applications}

End user applications like complaint registration systems, header, consent and content registration systems which need to interact with the blockchain system can continue to be built for users as traditional web applications by developers by reading the default key value data store provided by fabric or using a more performant mirror database for the global state of the blockchain. These end user applications can additionally also provide REST APIs or SDKs for their clients to interact with the blockchain using their node as a proxy. Self help portals provided by telecom operators can expose necessary APIs for their phone based and SMS based customer service systems to record complaints \& register customer preferences.

The democratization of the scrubbing service operation on the network prevents any vendor lock-in for smaller telemarketers and allows potential organizations to experiment with various pricing models based on the quality of service provided therefore creating a healthy, fair and auditable ecosystem. A key goal for the introduction of scrubbing service providers is to decouple trust placed in the service. The ecosystem of scrubbing providers, and auditors on the network can verify the operations thereby incentivizing, and creating an ecosystem for provable operational compliance.

\subsection{Automated complaint redressal \& tracking}
\label{ssec:complaint_redressal}

The availability of customer preferences and consents at any given snapshot, the content of the message sent, the customer phone number and the name of their telecom operator can be used to identify a transaction proposed by the telemarketer on behalf of a principal entity along with the proof of a scrubbing with the scrubbing token. This information can be replayed by observer/auditor nodes who register the complaint for validity. Using this approach, it becomes easy to find the participants who behaved maliciously and trace their actions which will be followed by a legal scrutiny. However, in a majority honest blockchain system such a situation would never arise because the encoded smart contract takes into effect these situations. If the situation still arises, it is due to a bug in the smart contract code and an updated policy with the contract can be issued to the blockchain for further usage by all the participants in the network with such a change being formally recorded.

However, Complaints against unregistered telemarketers making promotional calls without consent using previously leaked phone number data sets results in a complaint being registered with the maintenance of the watch list i.e. number of complaints registered against a specific number. On reaching a threshold number of violations, the observer nodes tracking the complaints can issue a special transaction requesting the telecom operators to provide the specific user degraded service. The telecom operators can then limit the total number of incoming or outgoing phone calls, SMS and access to mobile data of the violating subscriber behaving as an unregistered telemarketer and decide to terminate the phone number.

\section{Deployment Challenges}
\label{sec:deployment}

With the regulation draft finalized after the demonstrated feasibility of the systems and released on 18 July 2018, the regulators set out with the ambitious goal to enforce the new requirements starting Late December 2018 - Early January 2019. However, this received a lot of push back from the Cellular Operators Association of India (COAI) seeking the regulators pay attention to cost of implementation in the already financially-stressed telecom ecosystem indicating the technological overhaul to the current systems in place which become necessary due to the new mandate~\cite{comments-coai-2018,coaitotrai}. Additionally the COAI rebut the proposed regulation indicating potential migration of users to over the top (OTT) services like WhatsApp,  making the solution redundant after significant investments~\cite{coaitotrai}. However, COAI eventually realized that (1) they could not block the regulation through legal challenges, (2) the telecom operators could work together to reduce costs, and (3) the inclusion of a small scrubbing charge could defray the expenses.

The adoption of new technological architecture posed significant risk of service disruption to the telecommunications operators in India who eventually decided for a phased roll out of the infrastructure and developed a ``code of practice'' (CoP). The CoP describes how existing business operations would be transformed due to the new blockchain architecture, and helped establish a common procedure to be encoded into a smart contract for programmatic enforcement, essentially mandating all telecommunications operators to coordinate. In addition, the adoption of these systems faced 6 legal challenges since 2020 with cases filed in the Delhi and Karnataka High Courts, as well as the Supreme Court of India, by various telemarketers, and principal entities. A Public Interest Litigation (PIL) filed by advocate Reepak Kansal in the Supreme Court of India (\textit{Kansal v. The Union of India}) incorrectly argued to scrap the TCCCPR regulations on account of violating citizens' fundamental right to privacy due to the system creating a ``priceless, mega database of commercial relationships'' of over a billion individuals~\cite{pil_sc}. A 3 judge bench of the supreme court after reviewing the request refused to entertain the PIL request on 28th September 2020. Additional cases filed by a large mobile wallet and banking provider (One97 Communications, parent company of PayTM) used the updated TCCCPR'18 regulation to sue Telecom operators for failing to curb fraud of their registered header due to the similar deceptive headers, supporting the need for enforcement of the updated regulation. Smaller telemarketers filed legal cases claiming the new regulation hurt small businesses citing scrubbing enforcement resulting in over 80\% drop in their message volume (\textit{Shivtel Communications v. The Union of India} [Writ Petition (Civil) 3519/2021] at the Delhi High Court).

Eventually, the TCCCPR'18 regulation went into regulatory effect on 28th February 2019. This was followed by the enforcement of header registrations over the established blockchain network and scrubbing in September 2020, and soon followed by the need for registered templates and its inclusion in the scrubbing process on 8th March 2021.

The regulatory mandate was eventually adopted and put to practice adhering to the telecom operators' code of practice and included header, and content template registrations for the principal entities consuming SMS services. In \S\ref{sec:evaluation} we present, various results from a longitudinal study of spam and SMS messaging in India, and argue the effectiveness of the implemented solution to operate at the scale of a growing telecom market like India, improved compliance of registered telemarketers reducing spam, while improving the security and privacy of subscribers across the country who benefit from the large scale deployment of the system.

Despite the efforts and quantifiable evidence presented in this paper from a longitudinal study of spam and complaints in the Indian telecom ecosystem, the problem of spam has not yet fully been addressed with increasing spam being reported from phone numbers outside the regulated advertising ecosystem. We detail the challenges faced and outline research challenges in future work presented in \S\ref{sec:limitations}, and \S\ref{sec:discussion}.

\section{System Evaluation}
\label{sec:evaluation}

The deployment of the infrastructure complying to the TCCCPR'18 regulation opened up new opportunities for businesses to enter the telecom ecosystem and provide services to existing telecom operators and telemarketers. While the ecosystem through their code of practice and deliberations agreed to scrub both transactional (header + template scrubbing), and promotional (transactional + preferences + consent)  messages, continue using Hyperledger Fabric in their production deployment, various services like header, template, consent, scrubbing services were implemented directly by telecom operators, or other companies providing services to the telecom operators. As of today, the blockchain network consists of 7 Fabric orderer nodes run by the individual telecom operators (Vodafone-Idea Limited [VIL], Bharat Sanchar Nigam Limited [BSNL], Mahanagar Telephone Nigam Limited [MTNL], Airtel, Reliance Jio, Quadrant Televentures Limited [QTL], and Tata telecommunications). The network uses the Hyperledger Fabric v2 running RAFT consensus protocol in its deployment~\cite{Androulaki:2018:HFD:3190508.3190538, ongaro2014search} with each operator running  their own certificate authority (Fabric CA) in the blockchain network~\cite{Androulaki:2018:HFD:3190508.3190538}. We present our results below from one large scrubbing service provider in the ecosystem, and present evidence of reducing complaints, and increased adherence to the regulation in India.

\subsection{Boostrapping: Header \& Template Registrations from Principal Entities}
\label{ssec:bootstrapping_principal_entities}

By January 2020, the TCCCPR'18 regulation was very well known and the intent understood within the telecom operators. However, the fragmented ecosystem of principal entities relied on API partners providing them SMS services or telemarketers. Many of these stakeholders were unaware of the nuances of the new approach and how it could possibly disrupt their businesses. Many of them also did not pay attention to the notices from the telecom operators to register themselves again. Despite the readiness of the blockchain network in September 2019, the lack of engagement from businesses resulted in the slow on-boarding of principal entities and their registered headers into the new architecture.

This can be seen in Figure~\ref{fig:scrubbing_success_rate} with the solid red line denoting the registration of businesses. The spike in registrations seen between April 2020 to August 2020 was due to the issuance of multiple \textit{sunset date} warnings and notifications to non-compliant businesses about the strict scrubbing operations to be in-effect in the future resulting in a possible drop in their SMS messaging traffic. This resulted in increased awareness among some businesses about the new regulations and anticipating future requirements for content template scrubbing also registered their content templates used for messaging obtaining a registered template ID. This is shown in Figure~\ref{fig:scrubbing_success_rate} with the solid gold line indicating an increase in the number of template registrations between April 2020 and October 2020. The proactive measures taken to prevent lookalike headers during the registration phase resulted in over 23890 headers being rejected during registration. Post enforcement of header scrubbing, over a year in operation, not even a single complaint citing fraud was registered against any registered headers.

\subsection{Effectiveness of Scrubbing Services}
\label{ssec:effectiveness_scrubbing}

As the number of registrations of principal entities reached their previously registered quantities effectively signifying the transition of businesses towards the new architecture, the registered header scrubbing rules were enforced. The enforcement of these rules on 1st September 2020 (indicated by the dotted vertical orange line), meant that only the registered telemarketers could send promotional messages through the official established advertising channel.

\begin{figure}
    \centering
    \includegraphics[width=\linewidth]{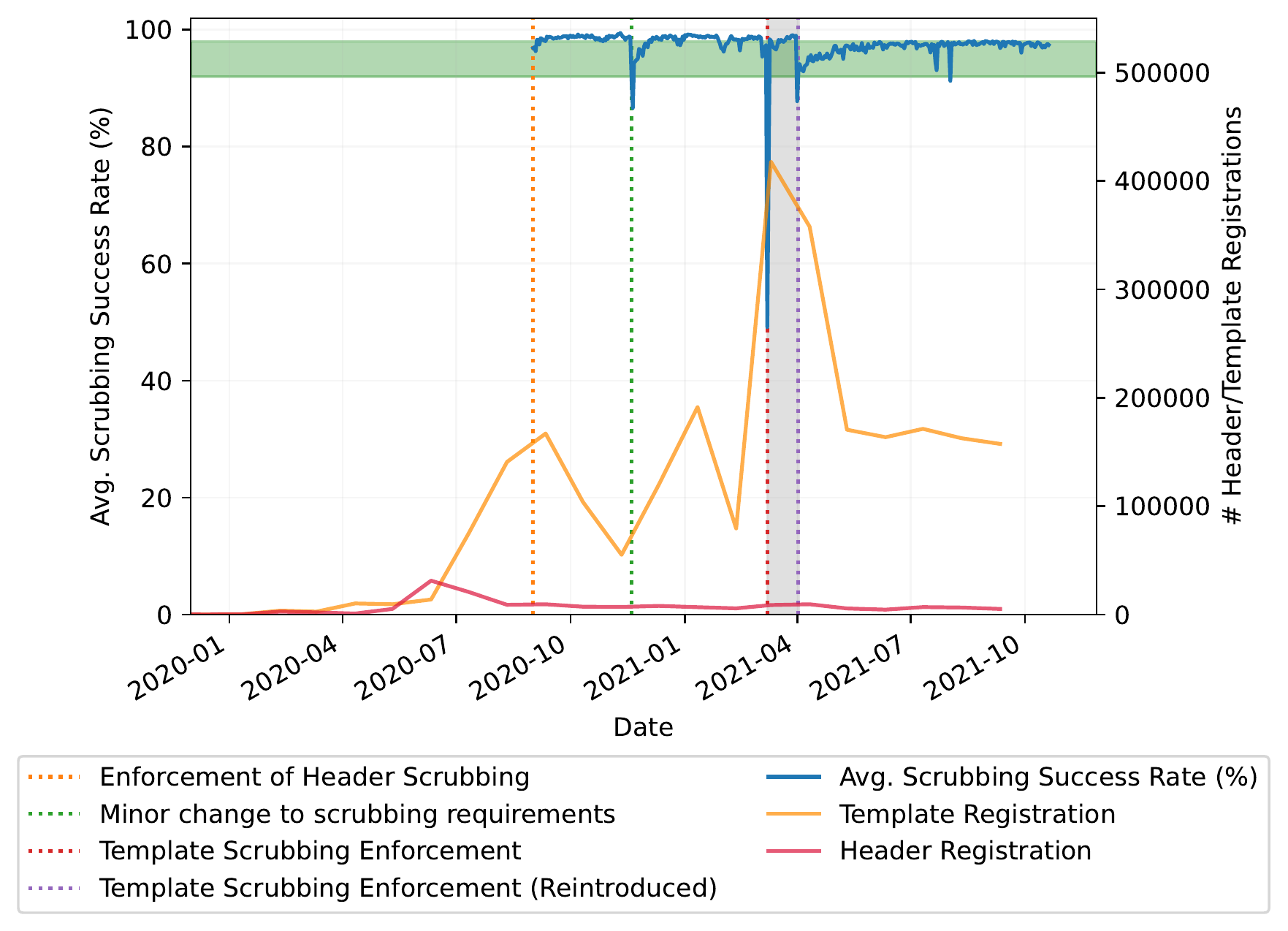}
    \caption{The figure shows the timeline of header (red) and template (yellow) registrations before the enforcement of the various phases of the regulations. The solid blue line indicates the scrubbing success rate and highlight various events resulting in a drop in the success rate. The green band shows the percentage success scrubbing events between 93 and 98\% respectively. The grey vertical band indicates the temporary relaxation of scrubbing constraints after a country wide drop in SMS traffic.}
    \label{fig:scrubbing_success_rate}
    \Description{Figure showing the timeline of various events overlayed with the header and template registrations seen by the network. The figure presents the average success rate of the scrubbing operation and shows a sharp drop in success after mandating template scrubbing.}
\end{figure}

We define successful scrubbing as the percentage of phone numbers from the submitted list of numbers, to which a successful message has been delivered. For example, a marketing campaign with 10000 phone numbers has a 99\% scrubbing success rate, if 9900 phone numbers provided by the telemarketers are in compliance and receive messages. The high scrubbing success rates could be for two reasons, (1) the default \textit{opt-in} model of promotional messages in the country could indicate majority numbers who have not exercised their DND preferences, and (2) the telemarketers have slowly become compliant and only send messages to phone numbers who have expressed an explicit preference, or are default \textit{opt-in} subscribers.

The new architecture and scrubbing requirements proved effective with an average over 95\% success rate in scrubbing on the first day gradually improving to over 98\% with compliant telemarketers removing the phone numbers from campaigns pro-actively. This is presented in Figure~\ref{fig:scrubbing_success_rate} as the solid blue line indicating the high number of successful scrubbing operations between September 2020 and November 2020. On 19th November 2020 (indicated by the dotted vertical green line), the operators updated their scrubbing requirements and operational workflows causing some message failures and degraded performance of the scrubbing system reducing the success rate to 85\%. However, this error was transient and was quickly addressed within a day resulting in over 95\% success rate. These results indicate the effectiveness of the scrubbing system in successfully identifying the 200M users in an ecosystem of a Billion exercising their preferences, and correctly preventing message delivery.

\begin{figure}
\hfill
\subfigure[The plot shows the increasing number of scrubbing operations being performed by one large scrubbing service provider managing the scrubbing operations for over 60\% of the registered businesses in the ecosystem]{
    \includegraphics[width=0.45\linewidth]{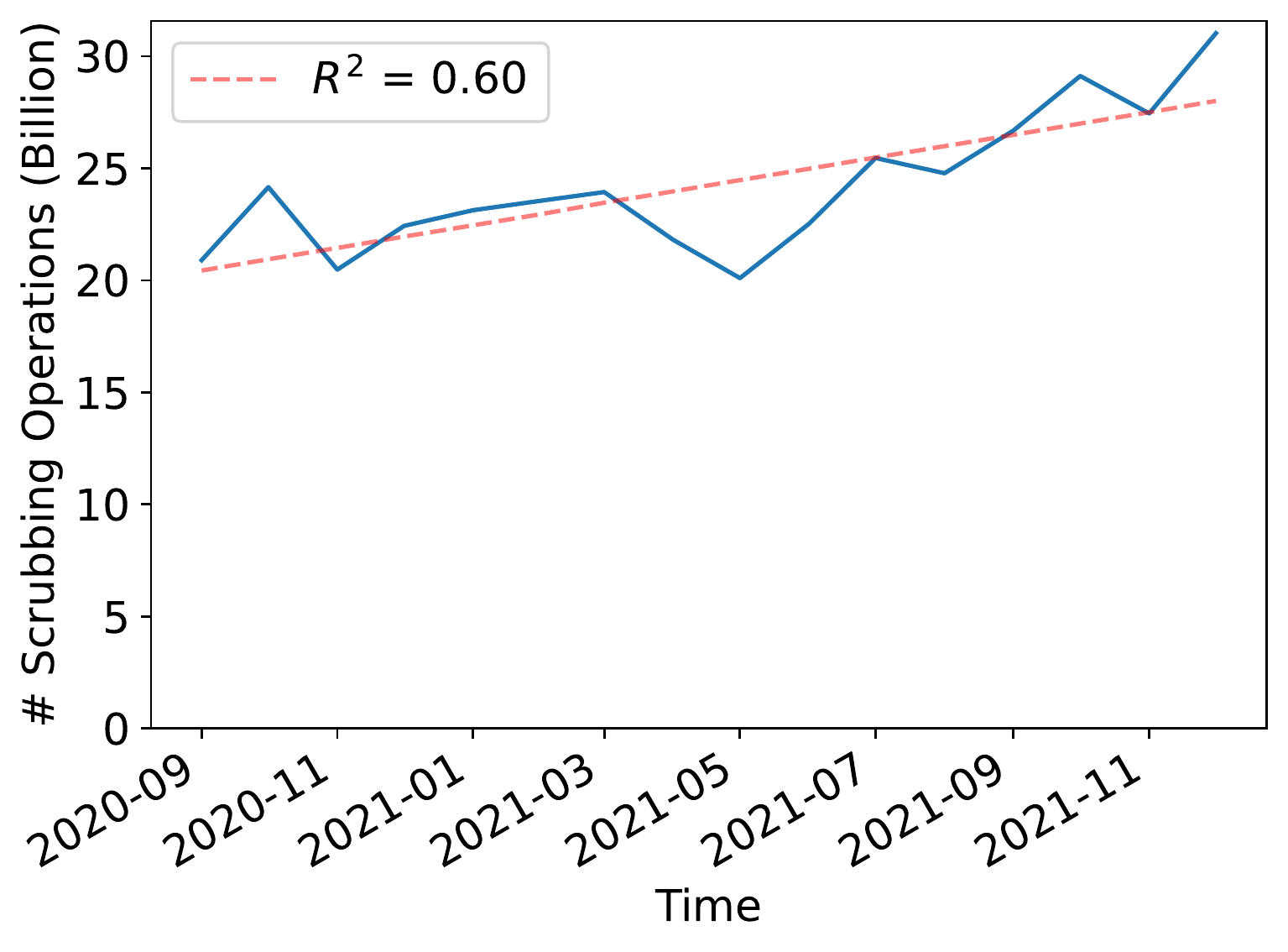}
    \label{fig:scrubbing_requests}
    \Description{The figure shows an increasing trend of the number of scrubbing operations being sent to the scrubbing services.}
}
\hfill
\subfigure[The plot indicates the total number of SMS messages being sent by one single provider and indicates an increasing trend and usage of SMS messages across the country.]{
    \includegraphics[width=0.45\linewidth]{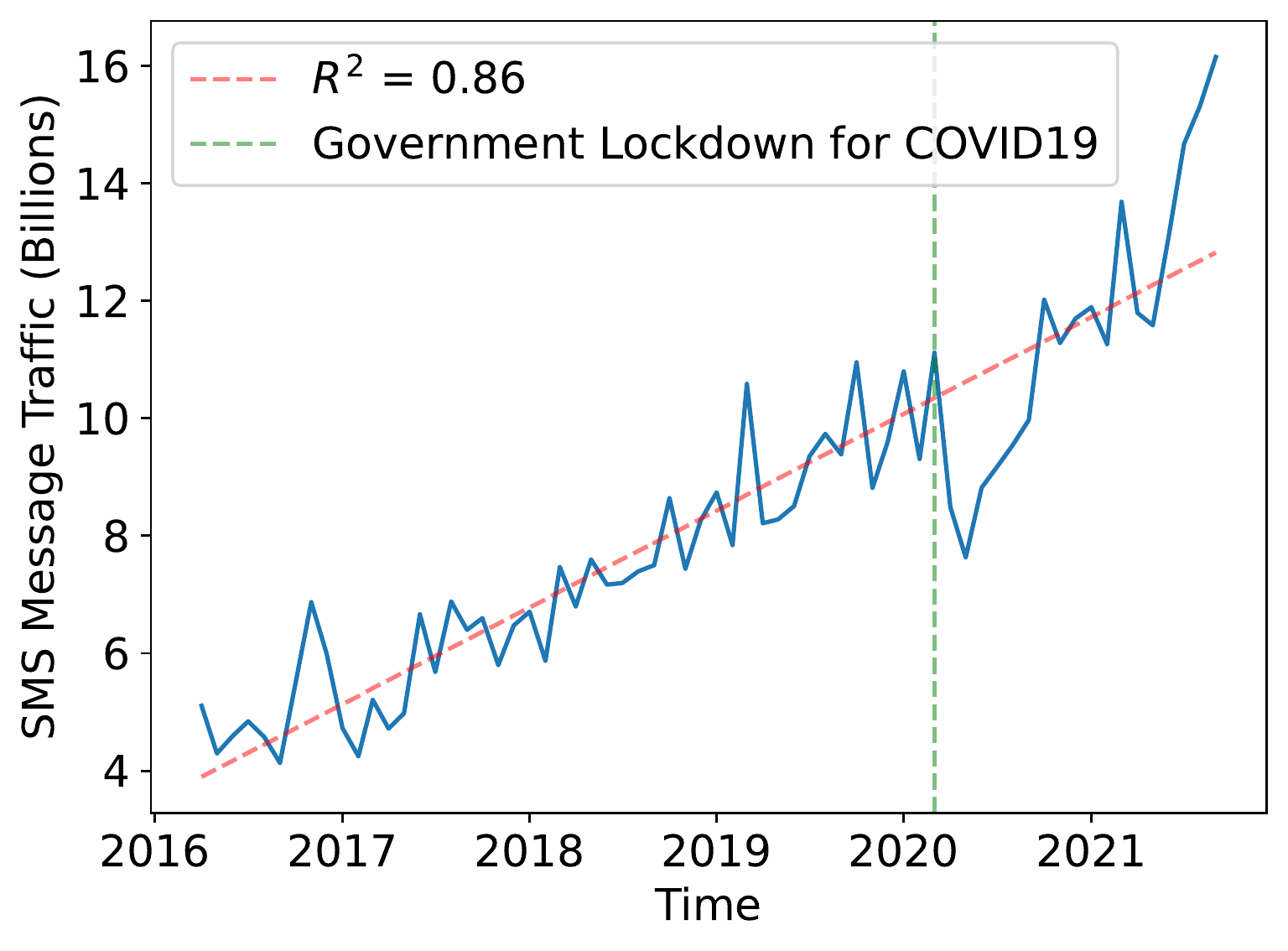}
    \label{fig:total_requests}
    \Description{The figure shows the total number of SMS messages being sent by one single provider indicating an increasing trend of usage of SMS across the country.}
}
\hfill
\caption{Volume of Scrubbing requests and SMS Messages Across India}
\end{figure}

The number of scrubbing operations to perform also increased over time with the scrubbing system currently handling 50\% more operations than it originally did when the system came into effect in September 2020, now handling over 30 Billion SMS scrubbing operations. The Figure~\ref{fig:scrubbing_requests} shows the increasing trend of scrubbing operations issued to the scrubber due to the increasing number of active SMSs being sent in the country (shown in Figure~\ref{fig:total_requests}). However, it is important to note that the measurement in Figure~\ref{fig:total_requests} only represents a partial view from one large operator and not the entire ecosystem. The overall ecosystem sends about a Billion SMS messages a day~\cite{lohchab2021}.

On 8th March 2021, the template scrubbing enforcement came into effect updating the scrubbing services to not only scrub verified headers and subscriber numbers but also validate the template of the message being sent. This resulted in an outage and a large drop in SMS delivery across the country with many critical banking, authentication, COVID-19 Vaccine registration information, and payment transaction messages being dropped resulting in only 49.2\% of messages being delivered after scrubbing~\cite{traidisrupt}. In addition to incorrect usage of template IDs in 41\% of the cases, and 5\% text mismatches between the registered templates, the major reason for the failure (54\%) was the lack of registration of templates by businesses due to possible miscommunication, or negligence. The disruptive nudge due to the drop of message traffic resulted in a spike in registrations as indicated by the orange line on March 2021 in Figure~\ref{fig:scrubbing_success_rate}. With the intervention of the regulator, the scrubbing process was relaxed for a week and eventually put into effect on 2nd April 2021 causing a minor disruption with drop in scrubbing success to 90\% before being returning to over 95\% success rate within a week.

The green bands show consistent successful scrubbing operations between 92\%-98\%. The overall mean success rate for scrubbing during the lifetime of the operations presented is $\mu$=97.5\%, standard deviation $\sigma$=3.01\%, and Median $M$=97.8\% indicating proactive efforts to curb spam, and improved operational practices by the telemarketers.

\subsection{Effect on Complaints Registered by Subscribers}
\label{ssec:effect_on_complaints}

\begin{figure}
    \centering
    \includegraphics[width=\linewidth]{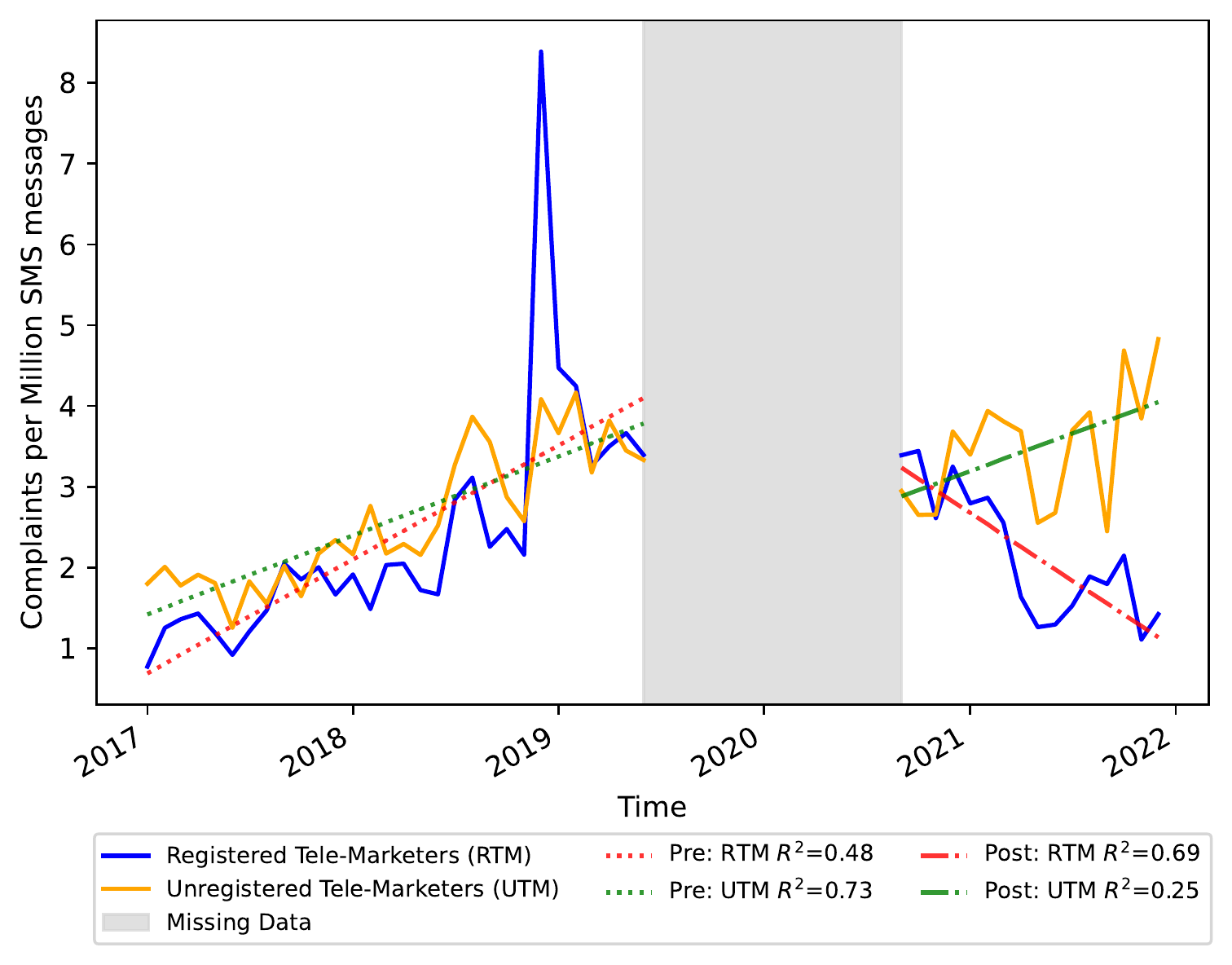}
    \caption{The plot shows a timeline of number of complaints against registered and unregistered telemarketers over time. The introduction of the blockchain system shows a decrease in the number of complaints against registered telemarketers while the number of complaints against unregistered telemarketers, and total complaints registered continue to rise.}
    \label{fig:complaint_rate}
    \Description{A timeline from 2017 - 2022 showing the increasing trend of number of complaints registered against RTM and UTM telemarketers from 2017 - mid 2019. The figure shows missing data between mid 2019 to December 2020 during the duration of the change in the production systems, The figure shows a decreasing trend in RTM complaints from Jan 2021 and sees an increasing trend in UTM complaints in the same period indicating the efficacy of the blockchain system for improved adherence to regulation and reduced complaints.}
\end{figure}

We use public records of the complaints registered by subscribers to TRAI through the app~\cite{androidtrai, iostrai}, or through their telecom operators and classify them into complaints against registered telemarketers (RTM) and unregistered telemarketers (UTM) based on the complaint against a specific registered header or against a valid 10 digit personal phone number through which unregistered telemarketing messaging has been carried out.

We present our results in Figure~\ref{fig:complaint_rate} and present the pre-blockchain and post-blockchain trends. The missing information between mid 2019 (development of the blockchain system), to mid 2020 (enforcement of mandates) is due to missing logs, and telemetry failures, resulting in data inaccuracies during the transition of systems. Historical data from 2017 to mid 2019 show an increasing number of complaints per million SMSs sent being registered. After the implementation of the blockchain based architecture and mandatory scrubbing, we notice a strong  reducing trend ($R^2$=0.69, indicated by the solid blue line) in the number of complaints against RTM with complaints down to 1.13 registered complaints per million SMS messages sent indicating the effectiveness of the system in reducing the mean complaints against RTM from 2.39 during mid 2019 indicating over 50\% improvement in the ecosystem, and increasing the marketers regulatory compliance. During the same phase, the number of complaints against UTM increased to 4.82 registered complaints per million messages sent (indicated by the solid orange line) making it easier for law enforcement authorities to identify spammers, fraudsters, and businesses violating subscriber consent.

However, the blockchain deployment phase, saw an increasing awareness among the subscribers who continued to report more complaints than in previous years. A majority ($>$75\%) of these complaints registered were against unregistered telemarketers resulting in such accounts used being banned, throttled, or blocked by the telecom operators. The reduction in the number of complaints against RTM despite the increasing number of overall registered complaints demonstrates the efficacy of the deployed system, and  its role in delivering the intended impact.

\section{Limitations}
\label{sec:limitations}

While the development, deployment, of the proposed solution, a year into its operation has shown promising results as shown in \S\ref{sec:evaluation}, the practical deployment of the blockchain system deviated slightly from its original proposal, and did not yet address its promise of curbing spam now increasing through unregistered telemarketing channels or due to voice calls. We outline some of the current limitations in our work below.

\subsection{Addressing Spam Calls and Unregistered Telemarketing Actions}
\label{ssec:limitations_calls}

Despite its ambitious inclusion in the TCCCPR'18 regulation to use the same technology proposal to address voice calls and completely curb UTM spam, the risk posed due to nation wide service outages and possible privacy violations resulted in the currently deployed solution completely detached from addressing robocalls (with automated voice, or human agent). While it is theoretically possible to perform content registration as spoken script templates in the current system, performing real-time checks of live calls is extremely difficult. It is also extremely hard to validate a complaint registered against a spam call without having access to the recordings of the conversations. Most telecom operators do not maintain this information but have provisions to explicitly target and do so for future calls to aid law enforcement actions when presented with a legal warrant. Performing these operations on every call is similar to wiretapping and severely compromises the privacy of subscriber communications since every phone call might potentially need to be monitored for identifying unregistered telemarketing calls initiated from malicious subscriber devices. The risks involved due to the voice data being biometric, and the severe impacts on privacy (intercepted SMS/voice information), implications on democracy, make this an extremely hard problem to tackle and open up opportunities for research on privacy preserving approaches to achieve these goals.

\subsection{Challenges with Delegations and Trust Relationships}
\label{ssec:limitations_blockchain_delegations}

The current deployment of the blockchain system in practice aims to have minimal disruption to principal entities carrying out their operations. Ideally, the principal entities would need to manage their cryptographic key identities and issue transactions (sending SMS / scrubbing) through the telemarketers. However, this would prevent various small businesses relying on SMS based marketing to gain experience and manage their keys. To simplify usage, the telemarketers act as delegated entities to the businesses and manage their keys, issue operations on their behalf. While the approach is a practical way to achieve the goals, it puts telemarketers at possible risk due to attacks from malicious adversaries wanting to compromise their systems and issue malicious transactions masquerading as the telemarketer affecting both the brand reputation of the principal entities and the telemarketers. The principal entities also in the process, continue to trust the telemarketers with plaintext phone numbers to send messages to which could be leaked to competing businesses by the telemarketer despite the telemarketer operating on hashed information when operating on the blockchain network. Scrubbing nodes are trusted not to misuse the phone numbers they receive from telemarketers in today's operations with active roadmap by the ecosystem to move towards reducing the trust placed in the service.

\subsection{Complexity, and Usability of Consent}
\label{ssec:complexity_consent}

The current blockchain system and scrubbing systems continue to enforce the seven high level categories and do not extend to the fine grained preferences and consent management allowing subscribers to exercise their explicit choices. Additionally, the workflows for recording digital consent, using them is incomplete and not standardized. Enabling fine grained preferences and clearly enabling subscribers to exercise their consent has user experience challenges and needs changes to various portals implemented by TRAI, Telecom operators, and mobile applications to be updated. However, the current system in place is highly extensible and can support these needs in the future. Additionally, the extension of the header registrations to full business names instead of 6 character alphanumeric codes has not yet been implemented.

\subsection{Translation of Code of Practices to Smart Contracts}
\label{ssec:translation_code_of_practices}

The code of practices established by the telecom operators identify various ways in which the telecom operators can inter-operate with other participants on the blockchain. However, the combined effort at translating the operational business logic into programmatic enforcements on the blockchain as smart contracts comes with the challenge of long deliberations among the participants before any changes are made which could impact business flow. Additionally, these smart contracts are agreed upon by the participants and have not yet received an audit for correctness of operations.

\section{Discussion}
\label{sec:discussion}

The proposed solution in this paper stems from the idea of allowing mobile subscribers complete control and transparency of their preferences. The solution introduces a procedure to manage and acquire consent digitally which can be used in more cases in the future such as granting consent to insurance companies to view health records, employers to access educational records etc.., The role of blockchain in Telechain is the first formal experiment with blockchain in the Indian government (also the first in the telecom industry worldwide), and is highly motivated with the idea of avoiding spam altogether by being compliant ex-ante instead of validating a complaint ex-post and holding an organization or entity to blame. The usage of blockchain in telecommunications in India which is rapidly growing is a chance for regulators to establish the potential of the technology to perform at scale and explore new areas where such technology could create a significant impact and avoid single points of corruptibility.

\subsection{\textbf{Strengthening Credibility, Collaboration \& Standardization}}

The usage of a blockchain while ensuring that the information can be accessed and read by anyone is a move to establish transparency which results in trust. By using blockchain and adopting the newly proposed systems the telecom operators in the telecom ecosystem can establish credibility by instantly taking into effect the subscriber preferences. The collaboration of the telecom operators of the country enabled via the network proves the ability of the technology to bring together mutually untrusted competitors, share necessary information and co-exist while still maintaining their business advantages. The usage of blockchains brings in the need for standardization and establishment of a common code of practice and cooperation between the entities.

\subsection{Cost and Benefits of the New Model}

While the creation of the blockchain network and securing it takes an investment of time and money, it also opens up new business opportunities. For example, We witnessed the growth of third party services monetizing services such as registration of headers, content and consent management, and providing scrubbing services. The ability to perform targeted reach to customers improves economic efficacy of telemarketers. The decentralized infrastructure improves adherence to regulations, makes participants responsible for each others' correct operations, and therefore prevents heavy penalties on the telecom operators and telemarketers while reducing the need for active regulatory oversight. The telecom operators and participants on the blockchain need to delegate dedicated infrastructure towards the network. We believe the costs of the infrastructure will be offset by the new business opportunities provided by improved quality of service and targeted reach.

\subsection{Ever growing data sizes \& Data storage}

The tamper proof and immutable nature of the blockchain means that the data continuously grows with time. The usage of data storage containers outside the blockchain to store paths to the previously scrubbed results results in increasing data storage costs due to space requirements. This information serves as valuable archival evidence for any future issues but can never be deleted without destroying the integrity of the blockchain.

\subsection{Practical Security of Blockchain Systems}

Blockchains enable mutually untrusted parties to collaborate and work together in a similar fashion. Blockchains if implemented correctly result in secure systems but there are various points where data leaks on the system can turn into security challenges. Any leak of keys stored in the membership services will result in malicious entities proposing faulty transactions. Similarly, colluding participants who can establish a majority in the network can tamper data and force the honest participants out of the network. While this can be counteracted with a policy stating that every single participant should validate the same content, it has an adverse impact on the performance and throughput of the system. The greatest risk to the consortium blockchain is posed by malicious system administrators who have privileges to the network and modify the network settings or tamper the keys maintained on the machine. However, these can be avoided by establishing an automated, verifiable and auditable release process. Consortium blockchains prove to be efficient in cases where mutually untrusted participants come together with a neutral regulatory oversight or auditor.

\subsection{Extensibility to Spam Calls, and Mobile Number Portability}

While the current proposed system focuses on curbing the issues due to unsolicited text messages, the same components on the blockchain can be used for tackling unsolicited phone calls. The TCCCPR'18 regulation takes into account the usage of the same architecture to curb spam phone calls. To avoid providing the scrubbed phone numbers to telemarketers, the scrubbing services provide aliased phone numbers which can be accessed via API interfaces on the blockchain enabling call based marketers to initiate calls to scrubbed numbers. This is done by initiating a Voice over IP (VoIP) based session with the telecom operator and providing the necessary scrubbing tokens received during scrubbing with authentication. The telecom operator validates and initiates the phone call to the actual number of the subscriber from the aliased value in the scrubbed list. Enabling such a solution would involve the 400-500 registered telemarketers in the country making phone calls to establish VoIP based SIP trunks with their partner telecom operators, update software to look up and initiate a connection than using an aliased phone number. By integrating such an approach, the drawback however is that the telemarketing agent initiating the call would not be able to look up the necessary information of the person being called without reconfirming their phone number and authenticating its match to the aliased number using APIs provided by the blockchain services. Implementing this process might extend the duration of the telemarketing call but help curb the problem of automated dialing calls or robocalls in the country.

\subsection{Learning from Deployment}

While the implementation of the proposed system, with the support of the ecosystem and driven by regulatory mandate resulted in partially achieving the goals we ambitiously set out with, it becomes evident that the adoption of the technology improves various high latency business operations involving synchronization and bring them towards real-time or near real-time latencies. Updates to subscriber consent now happens within minutes, and are used for scrubbing operations essentially reducing the time for consumer preferences to be in effect from 7-14 days in the past to within an hour and more ambitiously within minutes indicating an order of magnitude improvement in national scale operations. Additionally various fears from telecom operators about the loss of potential business posed by strict scrubbing requirements to over the top app channels such as WhatsApp, Telegram, etc.., remain unfounded due to the increasing volume of SMS as shown in Figure~\ref{fig:total_requests}.

The usage of the blockchain as a shared ledger allowed for more efficient tracking of complaints, with better information sharing established between telecom operators to identify the reasons for fraud and take immediate corrective action. This has resulted in uncovering fraudulent registrations by a few businesses with different telecom operators. With data sharing practices enabled through the blockchain, the different telecom operators could identify businesses registered under different names and carry out fraudulent activities. While the results presented indicate that the problem of spam is reducing through official advertising channels (Application to Person messaging), it is slowly moving towards a peer to peer model and being carried out by telemarketers masquerading as regular phone subscribers. The movement of spam to peer to peer channels makes it extremely difficult to identify the originator with current techniques without breaching the privacy and reading all the messages being sent by the subscribers. A recent survey by \textit{LocalCircles} indicated that 74\% of the respondents continue to receive unwanted SMSs despite being registered on the NDND registry with 26\% of them claiming these messages originate from service providers, or offers to earn more money~\cite{ec-localcircles-spam}. Future efforts at privacy preserving techniques to identify spam need to be extensively tested, deployed, and improved through iteration.

The changes to the NDND registry and hashing the necessary information resulted in malicious actors either performing rainbow table attacks or performing a brute force sweep of phone numbers given known telecom operator and regional prefixes but can easily be avoided by using keyed hashes~\cite{turner2008keyed}. It becomes necessary to deploy \textit{honeypot} solutions which detect such attackers, and filter them pro-actively amongst billions of legitimate subscribers without infringing on the privacy of the subscriber.

\section{Conclusions \& Future Work}
\label{sec:conclusion}

In this paper we look into the history of regulatory and technological initiatives taken to combat the problem of unsolicited commercial communication in the telecommunication industry. We detail the challenges in the current business processes adopted by  stakeholders in the ecosystem and propose a new consortium blockchain based architecture intended to address the problem of unsolicited commercial communication. We propose a solution using blockchain architecture that improves  subscriber experiences, regulatory compliance and also creates a vibrant ecosystem with economic opportunities for potential enterprises to operate in the telecom industry. The proposed solution has been incorporated into the latest regulation for UCC (TCCCPR'18) in India announced on \textit{19 July 2018}.

The solution has proven to be effective and operational at the scale of India, the second largest telecom market in the world with over a Billion active subscribers and with a message flow of over a Billion daily SMSs. The solution adequately met the design goals improving the speed, and efficiency of telecom operations, in addition to security, and privacy by preventing the sharing of plaintext subscriber information over the network. The implementation and enforcement of various suggestions in the proposal have significantly resulted in reduction of spam from organized advertising channels but see the flip effect of spam moving towards unorganized and peer to peer unregistered telemarketing practices needing further innovations. The establishment of the blockchain network enables new use cases such as mobile number portability and could also be applied to several other telecom services.

While present in the regulatory mandate, the current work does not yet establish a national consent registry allowing subscribers to exercise their ability to provide, and retract consent for advertising. However, the realtime nature of the new infrastructure will ensure that the consent options when introduced will take immediate effect.
Alternate efforts such as the focus on STIR/SHAKEN protocols improve identity and authenticity of users in telecom networks and could complement the efforts taken in this work towards identifying unregistered telemarketers masquerading as individual subscribers~\cite{mceachern2019shut, edwards2020robocalling, yu2021analysis}.
We leave this as future work for subsequent versions of \textit{Telechain}. The use of consortium blockchains with regulatory oversight is a promising way for governments to approach operational challenges involving mutually untrusted participants and enable cooperation to simplify and streamline their business operations while ensuring regulatory compliance. The effectivness of the solution described in this work inspired other countries such as the United Arab Emirates (UAE) to adopt the same blockchain based approach to curb spam in the UAE.


\begin{acks}

We are incredibly grateful to the deep commitment and efforts put in by various individuals who made this work possible, including telecom operators, individuals, organizations, and various stakeholders who engaged actively with the consultation papers. We thank all the stakeholders for their active feedback during the design, development, deployment of the infrastructure and Telecom Regulatory Authority of India for enabling these discussions. We'd like to thank various teams from Microsoft Research India, Microsoft Global Delivery, TechMahindra, and Baohua Yang from the Linux Foundation Hyperledger for their support during the development of prototypes. We'd like to acknowledge the valuable feedback provided on the paper by the anonymous reviewers, Waylon Brunette, Matthew Johnson, Miranda Wei, Ananditha Raghunath, and members of the ICTD Lab at the University of Washington.

\end{acks}
\bibliographystyle{ACM-Reference-Format}
\bibliography{referredpapers}

\end{document}